# Retention time trajectory matching for target compound peak identification in chromatographic analysis

Wenzhe Zang[†‡], Ruchi Sharma[†‡], Maxwell Wei-Hao Li[†‡§], and Xudong Fan[†‡]*

[†] Department of Biomedical Engineering
University of Michigan,
1101 Beal Avenue, Ann Arbor, Michigan 48109, United States

[‡] Center for Wireless Integrated MicroSensing and Systems (WIMS$^2$)
University of Michigan
Ann Arbor, Michigan 48109, United States

[§] Department of Electrical Engineering and Computer Science
University of Michigan
Ann Arbor, Michigan 48109, United States

*Corresponding author: xsfan@umich.edu

**Abstract**

Retention time drift caused by fluctuations in physical factors such as temperature ramping rate and carrier gas flow rate is ubiquitous in chromatographic measurements. Proper peak identification and alignment across different chromatograms is critical prior to any subsequent analysis. This work introduces a peak identification method called retention time trajectory (RTT) matching, which uses chromatographic retention times as the only input and identifies peaks associated with any subset of a predefined set of target compounds. RTT matching is also capable of reporting interferents. An RTT is a 2-dimensional (2D) curve formed uniquely by the retention times of the chromatographic peaks. The RTTs obtained from the chromatogram of a test sample and of pre-characterized library are matched and statistically compared. The best matched pair implies identification. Unlike most existing peak alignment methods, no mathematical warping or transformations are involved. Based on the experimentally characterized RTT, an RTT hybridization method is developed to rapidly generate more RTTs without performing actual time-consuming chromatographic measurements. This enables successful identification even for chromatograms with serious retention time drift. Experimentally obtained gas chromatograms and publicly available fruit metabolomics liquid chromatograms are used to generate over two trillions of tests that validate the proposed method, demonstrating real-time peak/interferent identification.

## 1. Introduction

Gas chromatography (GC)-based volatile organic compound (VOC) analysis can be classified into untargeted analysis and targeted analysis. The former involves evaluation of chemical substances in an unknown sample, whereas the latter aims only at a predetermined list of interesting compounds or a subset of those, with all other VOCs treated as interferents. Due to the complexity of sample composition and the lack of pre-existing knowledge, accurate identification in untargeted analysis requires confirmation or cross-validation by at least two parameters, such as chromatographic retention time (RT) and mass spectrometry (MS) fragmentation profile. In contrast, in targeted analysis oftentimes only the retention time is used for compound identification in order to avoid using bulky and expensive mass spectrometry. Therefore, targeted analysis has broad applications in on-site real-time measurements, such as environmental protection[1–3], working place environment monitoring[4–8], industries (*e.g.*, petroleum[9–11] and food[12,13]), and metabolomics[14–16].

For targeted analysis, the retention time of each peak in the GC chromatogram is compared with the pre-installed values of all compounds of interest in a library. In any given sample, a positive alarm is reported when the retention time of a peak matches a corresponding time in the library; the lack of any match instead means that a peak would be ignored or reported as an interferent. However, variations in physical factors such as ambient temperature, column temperature ramping profile, and carrier gas flow rate can affect the retention time of each peak from run to run, which hinders identification or triggers false alarms. The inability of correct peak identification with only retention times exacerbates when a sample contains a large number of compounds or some of the targeted peaks are closely eluted out in a chromatogram. Consequently, proper matching or alignment of chromatographic peaks across different samples is a crucial preprocessing step prior to any subsequent analysis.

A simple and popular solution is data binning, which divides the signals into bins (*e.g.*, histogram) and incorporate all data into a recognition profile for each measurement[17]. The binning method is easy to use and shows acceptable performance in processing both chromatogram and spectrum when the peak drift from sample to sample is much smaller than the distance between two adjacent peaks. However, in the presence of large peak drifts, this approach suffers from reduced resolution and information loss (illustrated in Figure S1). The time warping technique, such as segment-wise correlation optimized warping (COW)[18], point-wise dynamic time warping (DTW)[19], global polynomial model-based parametric time warping (PTW)[20, 21], multiscale peak alignment (MSPA)[22], and other variants[23,24], is one of most commonly adopted methods to correct retention time drifts across chromatograms. It aligns a whole measured chromatogram profile against a reference chromatogram using pattern recognition routines in order to achieve peak identification. While time warping is powerful and works well with samples of various complexities, an accurate warping-based aligning demands fine tuning of alignment parameters that can often involves human intervention, thus making automated peak identification less reliable. Moreover, all warping-based methods can suffer, to different degrees, from misalignments and are concentration-sensitive, even with samples of the same compositions. In some cases, warping-based aligning approaches may not be able to yield exactly the same retention time value for the same analyte from different measurements. Consequently, subsequent RT value based peak identification or statistical analysis often require further value correction via data binning or clustering. Machine learning based aligning approaches, which utilize artifical intelligence systems to acquire knowledge by extracting patterns from data, are also able to achieve positive alignment with decent accuracy[25–27], and they are amenable to automation without relying on human intervention. However, these approaches often employ the mass spectra of the peaks as one the of key subnetworks among the overall network achitechture, making it more suitable for bulky mass spectrometry based analysis rather than onsite monitoring. Also, similar to most machine learning based approaches, it suffers from high computational cost during parameter training and feature extraction.

While the aforementioned methods can mitigate the peak drift issue across chromatograms, they all suffer from a number of drawbacks (some of which are illustrated in Tables S4 and S5 in the Supplementary Information), particularly relevant to targeted analysis, which makes them unsuitable for compound identification in targeted analysis. First and foremost, the total number of peaks in the chromatogram obtained from the test sample needs to be exactly the same as in the reference chromatogram. If only a subset, or a single species in an extreme case, of the compounds of interest (target compounds), are present in the sample chromatogram, which is often the case for targeted analysis (*e.g.*, pipeline leakage detection in a chemical plant), chromatogram aligning fails, as the corresponding peaks have nothing to be aligned with (see examples in Figure S8 and Table S4A). Second, foreign interferents cannot be filtered out. If an additional peak is present in a chromatogram, it would be treated as one of the target compounds (misalignment) and/or it may cause failure in alignment of the whole chromatogram profile (see examples in Figure S9 and Table S4B). Consequently, there has been an unmet need for an MS-free and chromatogram-based peak identification method that is able to identify any arbitrary subset of the target compounds as well as to report interferents.

This paper proposes a retention time trajectory (RTT) matching method (Figure 1A) for peak identification. With peak retention times as the only input, RTT matching can identify peaks associated with any subset of target compounds as well as filter interferents outside of said targets. An RTT (Figure 1B) is made up of a series of retention times of all compounds (peaks) in a chromatogram obtained under one set of experimental conditions (such as ambient temperature, column temperature ramping profile, and carrier gas flow rate, *etc*.) and uniquely represents one particular condition. Similarities between sample RTT ($RTT_{sample}$) and those pre-installed ones in the library ($RTT_{lib}$) are globally evaluated using a statistical expression, until an $RTT_{sample}$ that best matches an $RTT_{lib}$s is found. Compared with most existing MS-free chromatogram aligning algorithms, peaks are identified via simple matching instead of chromatogram aligning. No mathematical warping or transformations are involved in the proposed method.

## 2. Retention time trajectory matching

### 2.1. Overview

Between GC analytical runs, the RT for a given compound may drift due to perturbation of various physical factors including ambient temperature, column temperature programing profile, and carrier gas flow rate. The influence of these perturbations on the analytes in a sample can be quite different due to their diverse characteristics (such as volatility, polarity, functional groups, *etc*.). Consequently, the RT drifts of the analytes in a chromatogram are often non-linear and unpredictable[20,28]. The RT deviation (ΔRT) against RT has been described using quadratic functions in PTW[20] or local regression fitting (LOESS) in XCMS[28], which often over-simplifies the diverse and complex nature of RT drifting. These methods are either limited to samples with the same constitutions[20] or require MS-based peak matching before final aligning[28]. In contrast, our RTT matching approach treats the RT drifts of *all* analytes (or peaks) in a chromatogram as a whole cohesive entity rather than independent individuals. Instead of fitting the RT drift with a mathematical formula, our RTT matching approach statistically compares the similarities between the $RTT_{sample}$s and the $RTT_{lib}$s.

First, we describe how to construct a library of $RTT_{lib}$s. Assuming that we obtain multiple chromatograms under various experimental conditions and each chromatogram includes *all* compounds, *i.e.*, target compounds and internal standard compounds (if needed). The X-axis of Figures 1B-D represents the retention time ($RT_X$) obtained from one of the measured chromatograms, which we call Chromatogram X. The dots of different colors along the X-axis represent different compounds. RTs of any pre-characterized chromatogram can be used as X-axis. Similarly, the Y-axis represents the retention time, $RT_Y$, in another chromatogram. Therefore, the set of coordinates, ($RT_{X,compound\ i}$, $RT_{Y,compound\ i}$), where "compound *i*" refers to a specific compound, form a trajectory in a 2-dimensional (2D) diagram (Figure 1B). Each trajectory corresponds uniquely to a chromatogram obtained under a certain set of conditions (temperature ramping and flow rate, *etc*.) and the entire sets of trajectories in the 2D diagram (*i.e.*, the library of $RTT_{lib}$s, Figure 1C) capture all chromatograms under various experimental conditions. One special case is that Chromatogram X itself is represented by a straight line 45$^o$ with respect to the X-axis (Figure 1D), as the coordinates along the X- and Y-axis exactly match each other.

Construction of an $RTT_{sample}$ for a test sample is similar (Figure 2A). Each $RTT_{sample}$ is made up of a series of coordinates, ($RT_{X,compound\ i}$, $RT_{sample,peak\ j}$), where $RT_{sample,\ peak\ j}$ refers to the retention time of a peak in the chromatogram obtained from the test sample (*i.e.*, sample chromatogram). Note that the peaks in the sample chromatogram may contain only a sub-set of target compounds (as well as interferents). Since the chemical identities of the detected peaks are unknown before analysis, multiple $RTT_{sample}$s may form for a given sample chromatogram, each of which corresponds to one set of peak identification results. Our goal is to eliminate all impossible $RTT_{sample}$s and find the one that best matches one of the $RTT_{lib}$s in the library.

In general, our RTT matching approach involves four steps:

(1) Experimentally generate multiple chromatograms under various experimental conditions (temperature ramping profiles and flow rate, *etc*.) using a mixture that contains all target compounds and internal standards (if needed). Retention times in each chromatogram are extracted, which form the $RTT_{lib}$s (see Figure 1C) pre-installed in the library. Note that the chemical identities of all peaks in any $RTT_{lib}$ are known.

(2) Create all possible $RTT_{sample}$s from the chromatogram obtained from the test sample (see Figure 2). Note that the test sample may contain only a sub-set of the target compounds and some interferents.

(3) Eliminate impossible $RTT_{sample}$s, which expedites computation.

(4) Compare all possible $RTT_{sampe}$s with the $RTT_{lib}$s in the library one by one. Find the $RTT_{sample}$ that best matches one of the $RTT_{lib}$s and then extract the chemical identities for the detected peaks accordingly.

Note that since an MS is not used, single compound injections should be conducted prior to Step (1) to estimate retention times and the relative elution order of the target compounds (and internal standards, if needed). Once this one-time task is completed, the RTs of target compounds and internal standards in each experimentally characterized chromatogram can simply be extracted based on the elution order.

In the next section, the following two cases will be considered. (1) Only a subset of compounds of interest (target compounds) and internal standards (if needed), are present in the sample chromatogram. (2) A few chemical interferents are present in a sample chromatogram.

### 2.2. Identification of a subset of analytes

Consider a sample with no interferents present (*i.e.*, all detected peaks are a subset of target compounds). We assume a total of $N_{tgt}$ target compounds and $N_{sample}$ detected peaks in the test sample. A matrix with $N_{tgt} \times N_{sample}$ intersections (coordinates) is then formed in the 2D diagram (marked as black dots, *i.e.*, coordinates, Figure 2A), since each peak in the test sample is unknown before final identification and in principle can be any of the $N_{tgt}$ target compounds. Consequently, a total of $\mathrm{C}(N_{tgt}, N_{sample})$ sets of $RTT_{sample}$s can be formed by connecting one black dot in each of the $N_{sample}$ rows, two of which are exemplified as black lines in Figure 2A. $\mathrm{C}(N_{tgt}, N_{sample}) = \frac{N_{tgt}!}{(N_{tgt}-N_{sample})!N_{sample}!}$ is a combinatorial number and can be an extremely large when $N_{tgt}$ is above 20 and $N_{sample}$ is around half of $N_{tgt}$ (for example, $C(20,10)$=184756).

However, not all $\mathrm{C}(N_{tgt}, N_{sample})$ $RTT_{sample}$s are possible and many need to be eliminated first before the comparison with the $RTT_{lib}$s in the library, which expedites computation and avoids false identifications. The elimination rules follow:

(1) One target compound can only be mapped to one peak in the sample chromatogram. Therefore, any $RTT_{sample}$ with a vertical section between any two consecutive coordinates (or intersections) in the 2D diagram should be eliminated, as illustrated in Figure S2A.

(2) The elution order should be preserved. Therefore, any $RTT_{sample}$ with a section that has a negative slope between two consecutive coordinates (*i.e.*, opposite elution order in library and sample chromatograms) should be eliminated. This is illustrated in Figure S2B.

(3) The RT drifts arise from minor perturbation and the resulting deviations (ΔRT) should be small values within a certain range. Therefore, only coordinates falling within the cutoff range of $RT_{sample}\pm\Delta t$ should be considered (Figure S2C). The value of Δt can be estimated empirically. For example, sample chromatograms with larger drifts require sufficiently large Δt (*e.g.*, larger than typical RT drifting range in the RTT library). Note that the RTT matching algorithm is still able to effectively identify the peaks even without applying this criterion, but a reasonable estimation of Δt significantly reduces the computational cost by narrowing down possible $RTT_{sample}$s.

Once all possible $RTT_{sample}$s are formed, each individual $RTT_{sample}$ should be compared with all $RTT_{lib}$s to find the best matched $RTT_{sample}$. In other words, we need to find which set of coordinates in Figure 2A fall on one red $RTT_{lib}$ in Figure 1C simultaneously. Assuming a total of $n_{lib}$ $RTT_{lib}$s stored in the library and $n_{sample}$ possible $RTT_{sample}$s generated from the sample chromatogram, $n_{lib} \times n_{sample}$ pairs of $RTT_{lib}$ and $RTT_{sample}$ are formed and compared with each other. For example, by comparing the trajectories in Figure 1B, two groups of black dots, which are composed of two different $RTT_{sample}$s, fall simultaneously on two different $RTT_{lib}$s (Figures 2B and C), respectively, and yield different chemical identification results for Peaks C and D. These are either Compounds 5 and 7 (Figure 2B), or 6 and 8 (Figure 2C).

To circumvent this, we introduce internal standard compounds (*i.e.*, internal standards) outside of the list of target compounds to anchor the $RTT_{lib}$s. In both $RTT_{lib}$ library preparation and actual measurement of the test sample, the internal standard(s) are spiked into the mixture containing all target compounds (for $RTT_{lib}$ library preparation) or the test sample. The peaks corresponding to these standards are identified during the data preprocessing and then used to generate the RTT along with all other peaks. As depicted in Figure 2D, when an internal standard (marked as a grey cross) is introduced, only one of the two $RTT_{lib}$s can be anchored (blue line) and thus a single set of identification result is obtained.

Introduction of internal standard(s) can further increase identification accuracy and significantly expedite computation by narrowing down possible $RTT_{sample}$s and $RTT_{lib}$s, since (1) All possible $RTT_{sample}$s and all $RTT_{lib}$s must go through the coordinate(s) formed by the internal standard(s). (2) Since the elution order is preserved, the whole 2D diagram can be divided into small regions determined by the internal standards' coordinates and only the $RTT_{sample}$s falling within these regions are possible candidates (Figure S3A). (3) When there is only a single analyte in the test sample, identification of its corresponding peak in the chromatogram is nearly impossible without internal standard(s), as formation of an RTT requires at least 2 coordinates. The addition of one or more coordinates resulting from the internal standards allows for the creation of the $RTT_{sample}$s for more accurate identification of the single peak. (4) Misidentification can be significantly reduced even when the chromatogram-to-chromatogram RT drift of the same compound is greater than the distance between two adjacent peaks (Figure S3B), which has long been the bottleneck of many peak matching or profile aligning algorithms[28]. In practice, internal standards can be strategically positioned in the region with more drastic variations to more effectively narrow down $RTT_{sample}$ and $RTT_{lib}$ selection (Figure S3C). Note, similar to all internal standards based chromatographic analysis methods, the addition of internal standards might potentially worsen the co-elution issues with the neighboring target compounds. To avoid this, internal standards whose RTs fall in the chromatogram sections with low peak densities are preferred.

Internal standards have commonly been utilized by many aligning algorithms[29,30], in which RT drift is corrected by first dividing the chromatogram into multiple sections delineated by the standards, and then applying linear stretching/compressing in each section  (Figure S4). However, these methods cannot account for the various non-linear drifts that often occur between any two standards. To make the linear stretching/compressing more accurate, more internal standards need to be introduced to reduce sections size to the point that linear approximation within each section is valid. This makes both sample preparation and peak identification much more complicated. More advanced techniques employ polynomial fitting within each section to account for non-linearity. However, polynomial fitting highly depends on experimental conditions and is required for each section in a chromatogram, thus hampering automation in peak identification. In contrast, as discussed in detail later, our approach compares the retention times of the chromatogram (or its corresponding RTT) globally, which automatically takes into account any non-linearities within each section. To demonstrate the advantages of RTT matching, we use the same sample and internal standards to compare the performance of the RTT matching approach and linear warping (see examples in Table S5) as well as correlation optimised warping (COW, see examples in Figures S8 and S9, and Table S4) in the Supplementary Information.

### 2.3. Statistical method – least mean squared residual (*MSR*)

As mentioned previously, while coordinates in the X-axis are discrete, their variation along the Y-axis is continuous due to continuously varying experimental conditions. In theory, the number of the $RTT_{lib}$s is infinite. In practice, only a limited number of conditions and hence a limited number of $RTT_{lib}$s can be characterized and stored. Consequently, while the $RTT_{sample}$ for a test sample may not exactly match any $RTT_{lib}$ stored in the library, the most similar one can still be easily found. To globally compare similarities between an $RTT_{sample}$ and an $RTT_{lib}$, we calculate the mean squared residuals (*MSR*) of RTs from the same compounds between these two trajectories, as illustrated in Figure 3. A smaller *MSR* indicates higher similarity between two trajectories, as exemplified in Figure 3B where $RTT_{lib(3)}$ is the most similar to the $RTT_{sample}$ in the figure. The *MSR* is normalized (or scaled) from the sum of squared residuals (*SSR*) by the total number of paired compounds to ensure that it does not grow as the number of pairs grows. This is important when we need to compare $RTT_{sample}$s with different numbers of target compounds, for example, an $RTT_{sample}$ of 6 target compounds is compared with an $RTT_{sample}$ of 5 target compounds plus 1 interferent (the case of interferent identification will be discussed later).

In order to expedite the computation, it is not necessary to compare each $RTT_{sample}$ with all $RTT_{lib}$s. Instead, we can first use internal standards RTs to anchor the best matching sets of $RTT_{lib}$s by calculating the *SSR* of internal standards ($SSR^{std}$), as shown in Figure 3A. Since all possible $RTT_{sample}$s must go through the coordinates formed by internal standards, they have the same $SSR^{std}$ for a given $RTT_{lib(i)}$, where $i$ refers to a specific pre-characterized $RTT_{lib}$. Assuming that there are a total of $N_{std}$ internal standards, the $SSR^{std}$ between any $RTT_{sample}$ and an $RTT_{lib(i)}$, denoted as $SSR_{lib(i)}^{std}$, can be calculated as

$$SSR_{lib(i)}^{std} = \sum_{k=1}^{N_{std}} \left( RT_{lib(i)}^{std(k)} - RT_{sample}^{std(k)} \right)^2, (1)$$

where $RT_{lib(i)}^{std(k)}$ is the retention time of one internal standard, $k$, in $RTT_{lib(i)}$, and $RT_{sample}^{std(k)}$ is the retention time of the same internal standard in the sample chromatogram. The $SSR^{std}$ of all $RTT_{lib}$s are sorted out ascendingly with the top ones, which have the least $SSR^{std}$s, giving the potentially matched $RTT_{lib}$s. All other $RTT_{lib}$s in the library, which have higher $SSR^{std}$s, can be eliminated.

Next, the $RTT_{sample}$ and $RTT_{lib}$ are further compared based on retention times of both internal standards and target compounds by sorting the *MSR* (Figure 3B). In this step, only the top $RTT_{lib}$s (*e.g.*, the first half or the top 20 $RTT_{lib}$s) with the lowest $SSR^{std}$s are selected. Assuming that, in addition to $N_{std}$ internal standards, there are $N_{sample}$ peaks to be identified in the test sample, the *MSR* between one $RTT_{lib}$ (denoted as $RTT_{lib(i)}$) and one $RTT_{sample}$ (denoted as $RTT_{sample(j)}$) can be calculated as

$$MSR = \frac{SSR_{lib(i),sample(j)}}{N_{std}+N_{sample}} =$$

$$(SSR_{lib(i)}^{std} + \sum_{l=1}^{N_{sample}} \left(RT_{lib(i)}^{compound(l)} - RT_{sample(j)}^{compound(l)}\right)^2)/(N_{std} + N_{sample}), \quad (2)$$

where $RT_{lib(j)}^{compound(l)}$ is the retention time of the Compound *l* in $RTT_{lib(i)}$, and $RT_{sample(i)}^{compound(l)}$ is the retention time of a peak in the sample chromatogram that is hypothetically assigned to the same compound (*i.e.*, Compound *l*) in $RTT_{sample(j)}$. Note that all $RTT_{sample}$s have the same retention time for each peak, but each peak can hypothetically be paired with a different compound in different $RTT_{sample}$, which has been discussed previously (*e.g.*, Figure 2A). The *MSR* of all $RTT_{lib}$s are sorted out ascendingly. The first in the list has the minimum *MSR* value denoted as $MSR_{sample(j),\,min}$, which is generated by the $RTT_{lib}$ that best matches $RTT_{sample(j)}$.

Since each $RTT_{sample}$ is formed by pairing detected peaks with one set of target compounds, it represents one set of peak identification results. For any $RTT_{sample(j)}$, the best matched $RTT_{lib}$ can be found by screening all $RTT_{lib}$s in the library and finding the one that generates $MSR_{sample(j),\,min}$. If there is another $RTT_{sample}$, denoted as $RTT_{sample(j')}$, that has $MSR_{sample(j'),min}$ smaller than $MSR_{sample(j),min}$, it means that $RTT_{sample(j')}$ better matches one of the $RTT_{lib}$s in the library. Therefore, corresponding identification results from $RTT_{sample(j')}$ are more possible than those from $RTT_{sample(j)}$.

A glossary of abbreviations and symbols is summarized in Section S5 in the Supplementary Information.

## 2.4. Interferents

In targeted analysis, interferents are the compounds not on the list of target compounds and need to be filtered out. In our algorithm, two criteria are used to identify the presence of interferents. First, for a particular peak in the sample chromatogram, if none of the retention time values in the $RTT_{lib}$s falls in the range of $RT_{peak} \pm \Delta t$, this peak is identified as an interferent. The value of $\Delta t$ can be chosen empirically and can be set higher than a typical retention time drift range. Second, for one particular pair of $RTT_{sample(i)}$ and $RTT_{lib(j)}$, if the squared residual between one peak (for example, Peak *D* in Figure 3C) in $RTT_{sample(i)}$ and its paired compound (Compound 9 in Figure 3C) in $RTT_{lib(j)}$ is much larger (*e.g.*, twice) than $MSR_{lib(i),sample(j)}$, it is highly likely that this peak is an interferent. The validity of this approach lies in the fact that all other coordinates formed by the detected peaks and their paired target compounds well match $RTT_{lib(j)}$, except for the one formed by Peak $D$ paired with Compound 9. A new $MSR_{lib(i),sample(j)}$ is then calculated by normalizing $SSR_{lib(i),sample(j)}$, which excludes residuals from all identified interferents, with $N_{std} + N_{sample} - N_{interf}$ ($N_{interf}$ is the total number of identified interferents). Based on this, all possible peak identification results, with or without interferents, can be ranked by *MSR*s. The results with the smallest *MSR* give the highest confidence level.

## 3. Experimental

### 3.1. Chromatogram generation

Our RTT approach is validated using nine chromatograms obtained with NovaTest P300 GC provided by Nanova Environmental, Inc., which is equipped with a 6 m long Rtx-VMS column (Restek, Bellefonte, PA, USA) and a microfluidic photoionization detector developed in-house[31]. The chromatograms were generated under the same nominal experimental setting (carrier gas: helium; flow rate: 3.5 mL/min; temperature programming profile: 40 °C held for 5 min, ramped to 70 °C at 30 °C/min, held for 2 min, then ramped to 150 °C at 30 °C/min, and held for 1 min). The injected mixture is part of EPA Method TO-14. Exemplary chromatograms are presented in Figure 4A and corresponding peak information is listed in Table S1.

### 3.2. Chromatogram preprocessing

Detection of a peak in a chromatogram is accomplished by scanning for local maxima and the associated peak apex positions (*i.e.*, retention times)[32]. A series of retention times are extracted, which are used to form the $RTT_{lib}$s or $RTT_{sample}$s in the next section. Therefore, the cumbersome chromatographic data (a large 2D array of detection signals) are converted to a simple list of retention times, which significantly reduces data storage and processing workload. Extensive preprocessing (*e.g.*, baseline removal) and broad background variations can also be eliminated since only the local maxima (*i.e.*, peak apexes) are extracted.

## 4. Validation results

Out of nine experimentally generated chromatograms, six chromatograms (denoted as $Chrom_{1-6}$) are used in the library, forming $RTT_{lib}$s (Figure 4B. The remaining three (denoted as $Chrom_{7-9}$) are used to generate tests to validate our approach in various scenarios.

Detailed validation tests design is described in Section S1 in the Supplementary Information. There are a total of 22 peaks in each measured chromatogram, among which 20 are treated as target compounds and the other two are used as the internal standards. The retention times and compound IDs of $Chrom_1$ are summarized in Table S1. The RT deviation (ΔRT) against the RT in $Chrom_1$ for all chromatograms ($Chrom_{1\text{-}9}$) is plotted in Figure S5, showing strong non-linear drifting behavior. Note that while the chemical names for most compounds are known, which are given in the third column in Table S1, the chemical identities of the first and the third eluted peaks are unknown (which might result from contamination and are only designated as ID 1 and ID 3, respectively). Nevertheless, the results presented in this work remain the same regardless of whether the chemical names of those compounds are known.

In total, $3 \times \sum_{i=1}^{20}[\mathrm{C}(20, i)] = 3.15 \times 10^5$ validation tests are generated from $Chrom_{7\text{-}9}$, covering *all* subsets of the 20 target compounds (ranging from single compounds to 20 compounds). Moreover, 3 additional validation tests are generated, representing samples with a subset of target compounds *and* interferent(s). In all validation tests, peak identifications achieve 100% accuracy. Among all the validation tests, we present detailed results of 11 representative tests, covering various MS-free chromatographic analysis scenarios.

**Scenario 1:** different levels of retention time (RT) drift

(1) The RT drift is within the RT drift range of the pre-characterized RTTs (see Tests 1, 2, 3 with $Chrom_7$, Table 1A and Figure S5).

(2) The RT drift is slightly out of the RT drift range of the pre-characterized RTTs (see Tests 4, 5, 6 with $Chrom_8$, Table 1B and Figure S5).

(3) The RT drift deviates far from the RT drift range of the pre-characterized RTTs (see Tests 10 and 11 with $Chrom_9$, Table S3)

**Scenario 2:** various sample components with different numbers of target compounds

See Tests 1-6, 10 and 11 (out of a total of 20 target compounds, the target compound number in each test is 11, 13, 5, 9, 5, and 1, respectively). Especially, single peak is successfully in Test 6.

**Scenario 3:** with or without interferents

Tests 7, 8, and 9 (Table S2) respectively represents samples of 5 target compounds plus a single interferent (interferent RT is far from neighboring target compounds), 5 target compounds plus single interferent (interferent is close to neighboring target compounds), and 9 target compounds plus 2 interferents. The remaining tests represent samples containing a subset of target compounds.

### 4.1. Identification of target compounds and interferents

To validate our algorithm, we first discuss the scenarios in which no interferents are present and all the detected peaks are a subset of the target compounds. Based on the RTs in $Chrom_7$ and $Chrom_8$, six groups of RTs are generated, three from $Chrom_7$ and three from $Chrom_8$, for six tests. These represent six different mixtures containing various subsets of the target compounds along with two internal standards ($std_1$ and $std_2$). Note that the retention time deviation in $Chrom_7$ is within the range of the RT deviations in the chromatograms ($Chrom_{1\text{-}6}$) stored in the library, whereas the retention time deviation in $Chrom_8$ is slightly out of this range (Figure S5). For each test, four best peak identification results are enumerated based on the *MSR* (Table 1). The peaks in all six tests are successfully identified with the top result (*i.e.*, smallest *MSR*) producing 100% accuracy. The $2^{nd}$-$4^{th}$ best results in each test also correctly identify most of the peaks with the best ones giving 100% accuracy and the worst ones misidentifying only one peak (marked with an asterisk "*"). Note that even a single-species sample (Compound 7 in Test 6) can be correctly identified due to the use of internal standards, despite it being very close to neighboring Compound 8. This is impossible for all warping-based chromatogram aligning approaches, since the peak has nothing to be aligned with.

Another three validation experiments (Tests 7-9 in Table S2) are generated based on the RTs in $Chrom_8$, which mimics scenarios in which both a subset of target analytes and interferents are present. In Tests 7 and 8, one hypothetical interferent peak is added at 340 s and 449 s, respectively. In particular, the added peak at 449 s is very close to the target Compound 19. In both cases, the top identification result successfully identifies all target compounds and singles out interferent related peaks with 100% accuracy. In Test 9, two hypothetical interferent peaks are added at 62 s and 395 s, respectively, which are very close to target Compounds 6 and 16. Both interferents and all target compounds are correctly identified. Like all other RT based peak identification methods discussed previously, the RTT matching based interferent identification works only when $RT_{interferent}$ is sufficiently different from those of target compounds. If $RT_{interferent}$ is the same as or extremely close to any target compound, the interferent cannot be identified. Additionally, the majority of the peaks in the sample should be the target compounds. If most peaks are interferents and only a few target compounds are present, the validity of our method may decrease, since the residuals of most peaks are very large. To circumvent these issues, one way is to further enrich the RTT library, either experimentally or through RTT hybridization. Introducing more internal standards might also be necessary.

It is also worth noting that the proposed RTT matching approach is intended for scenarios where RT drifts result from only *minor* fluctuations in experimental conditions and therefore the elution order is expected to hold among the measurements. When the experimental condition varies drastically (*e.g.*, major changes in the device settings or ambient temperatures), the elution order may vary. Therefore, a new RTT library needs to be constructed under the new experimental conditions to avoid misalignment/misidentification in the RTT matching approach.

The above two issues (*i.e.*, serious co-elution and elution order change) have always been the bottleneck for all existing MS-free chromatogram aligning algorithms. MS (and other spectroscopic methods such as infrared absorption spectroscopy) would potentially be needed for peak identification in these cases.

**4.2. RTT library optimization - "hybridization"**

An ideal RTT library should contain all possible $RTT_{lib}$s that cover all possible drift inducing conditions. If the library has only a limited number of $RTT_{lib}$s and when the sample chromatogram drift (or retention time deviation) exceeds the retention time deviations covered by the $RTT_{lib}$s, peak misidentification may occur, as exemplified by Tests 10 and 11 in Table S3. One method to enrich $RTT_{lib}$s is to experimentally generate as many $RTT_{lib}$s as possible by varying the experimental conditions around the nominal conditions. However, this is extremely labor intensive and difficult to realize. Alternatively, the new $RTT_{lib}$s can be generated by linearly hybridizing existing experimentally obtained $RTT_{lib}$s. This method is valid because the retention time drift is caused by minor fluctuations of system physical factors and such a small perturbation in the retention time of one particular compound from one state (one RT) to another state (another RT) can be simplified as linear variation. RTT linear hybridization can be done using two, three, or more existing experimentally obtained $RTT_{lib}$s, *i.e.*,

$C_1 \times RT_{lib(a)}^{compound(i)} + C_2 \times RT_{lib(b)}^{compound(i)}$, or

$C_1 \times RT_{lib(a)}^{compound(i)} + C_2 \times RT_{lib(b)}^{compound(i)} + C_3 \times RT_{lib(c)}^{compound(i)}$,

where $C_1$, $C_2$, and $C_3$ are the linear coefficients, and $RT_{lib(a)}^{compound(i)}$ refers to the retention time for Compound $i$ in one of the $RTT_{lib}$s (*i.e.*, $RTT_{lib(a)}$). The hybridization can easily generate more $RTT_{lib}$s of the intermediate states that may be difficult to obtain experimentally (due to either time limitations and/or difficulty in realizing the exact experimental conditions), which significantly increases the tolerance to more serious RT drifts. Note that the RT variation of one particular compound from one state to another is simplified to be linear, although ΔRT against RT in one chromatogram is generally non-linear.

To validate the hybridization method, in the current work we use two-RTT hybridization based on following three formulas: (1) $\left(RT_{lib(a)}^{compound(i)} + RT_{lib(b)}^{compound(i)}\right)/2$, (2) $2RT_{lib(a)}^{compound(i)} - RT_{lib(b)}^{compound(i)}$, and (3) $2RT_{lib(b)}^{compound(i)} - RT_{lib(a)}^{compound(i)}$ to enrich the $RTT_{lib}$s in the library, where $RTT_{lib(a)}$ and $RTT_{lib(b)}$ are from $Chrom_{1\text{-}6}$ that are experimentally generated. Two tests (Tests 10 and 11) are generated based on $Chrom_9$, which has much more serious drift compared to $Chrom_{7\text{-}8}$ and any pre-characterized chromatograms ($Chrom_{1\text{-}6}$) in the library (see Figure S5). As shown in Table S3, when the RTT library contains only the experimentally generated $RTT_{lib}$s, it fails to correctly identify all peaks in either Test 9 or 10. In contrast, with more $RTT_{lib}$s added through hybridization, all peaks are successfully identified with 100% accuracy. The accuracy of the $2^{nd}$-$4^{th}$ identification results in both tests are also increased.

### 4.3. Comparison with other chromatogram aligning approaches

In order to compare peak identification performance with other chromatogram aligning approaches, part of the above validation tests are also performed with whole chromatogram based COW[18,19] method and two peak-list-based aligning methods, namely internal standard based linear warping approach and fast PTW[21]. Herein, $Chrom_1$ (or its corresponding peak list) is treated as the reference to be aligned with.

Tests 5 and 7 (based on $Chrom_8$), which respectively represent a sample without and with an interferent are used to evaluate COW and linear warping. The retention times after COW or linear warping[30] are summarized in Tables S4 and S5, respectively. Note that in the reference chromatogram (*i.e.*, $Chrom_1$), all the peaks are present, including *all* target compounds and internal standards. The sample chromatograms, which are to be aligned and identified, are reconstructed from $Chrom_8$, but contain only the peaks listed in Tests 5 and 7. The remaining peaks in $Chrom_8$ are replaced with the baseline (see Section S2 in the Supplementary Information for details of sample chromatogram reconstruction).

The COW aligning results can be highly parameter dependent[18]. When the peak compositions in the reference and sample chromatograms are same, optimal parameters can be easily chosen so that the apex position of the peak of the same elution order coincides between the sample chromatogram and the reference chromatogram. However, in the presence of only a subset (as in Tests 5 and 7), the peak in the sample chromatogram has no specific target peak to align to, and therefore, the COW alignment parameter selection becomes dubious. Similarly, the interferent peak (as in Test 7) might be misaligned to one of the target compound peaks in the reference chromatogram. Multiple COW alignments with various tuning parameters (slack and correlation power) were conducted. The identification results based on the RTs extracted from the aligned $Chrom_{sample}$ are summarized in Table S4. Out of the 7 peaks in Test 5, the best COW aligning only correctly identifies 5 peaks (slack=4 and correlation power=3) and the worst aligning fails to align any of the peaks (slack=1 and correlation power=1; slack=2 and correlation power=2). For Test 7, the highest identification accuracy reaches only 50% (slack=4 and correlation power=3) and the worst aligning fails with the whole chromatogram (slack=1 and correlation power=1; slack=2 and correlation power=2).

For the internal standard based linear warping method (Table S5), the same internal standards ($std_1$ and $std_2$) are employed. After alignment, none of the target compound associated peaks yields the same RT as in the reference chromatogram ($Chrom_1$). Therefore, peak identification for target compounds fails when RT values are compared (all peaks are identified as interferents, except the interferent peak itself).

Additionally, we compare the peak identification performance of fast PTW algorithm[21], which is a further development of the original PTW[20] and the algorithm input is chromatogram peak list (retention times and peak heights). The peak listed in $Chrom_1$, which is treated as the reference, and $Chrom_{8,9}$, which are used to generate Tests 5, 7, 10 and, 11, are summarized in Table S6A. The time warping results (order = 2 or 3) using the full peak lists in $Chrom_{8,9}$ are summarized in Table S6A. Although misalignments are greatly reduced after fast PTW aligning, sub-second to seconds of RT difference for the same compound persists. Table S6B summarizes the fast PTW aligning results (order = 2 or 3) of Tests 5, 7, 10, and 11. Tests 5, 10, and 11 represent samples with only a subset of target compounds; Tests 7 represents a sample with a subset of target compounds and an interferent. The fast PTW fails in all four tests to identify correct peaks. In comparison with the reference, the RT differences of the same compound in the warped sample peak list fall in the range of sub-seconds (which could be solved by data clustering) to hundreds of seconds (which completely fails in peak identification). We also notice that the warped RTs of some peaks give an unreasonable negative values (*e.g.*, Compound 1 in Test 10 with order = 2, and Compound 17 in Test 10 with order = 3).

## 4.4. Further validation with fruit metabolomics data

Two publicly available liquid chromatography (LC) fruit metabolomics datasets from the Metabolights repository (http://www.ebi.ac.uk/metabolights) with identifiers MTBLS99 and MTBLS85 are used to generate additional validation tests. This demonstrates the application of the RTT matching algorithm to complicated samples and other chromatographic techniques. The first dataset consists of LC measurements of a pooled sample that was injected regularly as a quality control during the measurements of apple extracts (Figure S10A). The second datasets are from LC measurements of carotenoids in grape samples (Figure S10B). The test design and results are summarized in Section S4 in the Supplementary Information. In all of the 2 trillion designed validation tests, all target compounds associated peaks are correctly identified.

## 5. Discussion and conclusion

We have described in detail and validated a new RTT matching approach for identification of target compounds and detection of interferents. This approach has the following main features. First and foremost, the matching is conducted between the entire trajectories ($RTT_{sample}$ and $RTT_{lib}$) rather than between the individual peaks in the sample and the reference chromatogram. Second, simple statistics is used, which avoids time consuming training or feature extraction used in machine learning. The *MSR* is adopted to describe the similarities between RTTs, which works well with the chromatograms obtained in this article. Other statistical approaches, such as linear regression, can also be introduced within the framework of RTT matching. Third, hybridization of $RTT_{lib}$s greatly enriches the RTT library, which not only increases the tolerance to more serious drifting, but also significantly reduces the cost for $RTT_{lib}$ generation through actual experimentations. In this work, each hybridized $RTT_{lib}$ is generated out of two experimentally obtained $RTT_{lib}$s. In the future, a larger number of experimentally obtained $RTT_{lib}$s could be involved and the linear coefficients could be further tuned depending on the required complexities of analysis the subjects to be analyzed. Fourth, the only input variables are the retention times of each peak instead of the whole peak profile or mass spectra, making the RTT approach insensitive to concentration or background, and eliminating the need for bulky instruments such as mass spectroscopy. All of the above features make RTT matching highly amenable to automation with low computation cost. Additionally, the method described here can be easily translated to other chromatographic techniques (*e.g.*, ion exchange chromatography) and has great potential to be applied to other spectral data (*e.g.*, nuclear magnetic resonance spectroscopy). Finally, the introduction of internal standards, though contributing to increasing identification accuracy and computation efficiency, are not always necessary. When an increased number of target compounds are present in the sample, the contribution of standard(s) becomes less prominent. To validate this, 10 of the validation test examples in the above discussions, except the single-species sample in Test 6, have been re-performed with RTT matching without use of any internal standard. All the peaks can be correctly identified.

Finally, it is worth noting that the applications of the RTT matching approach are not limited to just peak identification. For visualization purposes, chromatogram aligning can easily be achieved with the RTT matching approach. Briefly, one can choose any $RTT_{lib}$ as the reference chromatogram to extract RTs of target compounds. Each peak in the sample chromatogram is identified via RTT matching approach. Peak profile can be fitted by the exponentially modified Gaussian (EMG) model with apexes shifted to the positions as in the reference. Aligned sample chromatograms can be formed by the summation of individual EMG reconstructed peaks. Detailed description and illustration is discussed in Section S4 and Figure S11.

## AUTHOR INFORMATION

**Corresponding Author**

* E-mail: xsfan@umich.edu

**Conflict of interest**

Xudong Fan is a paid consultant for Nanova Environmental Inc., whose portable GC system was used to generate the exemplary chromatograms analyzed and presented in this article.

## ACKNOWLEDGMENT

The authors acknowledge the support from National Institute for Occupational Safety and Health (NIOSH) via grant R01 OH011082-01A1, Beijing Institute for Collaborative Innovation via the University of Michigan internal funding, Go Blue Biotech via the University of Michigan internal funding, and the Office of the Director of National Intelligence (ODNI), Intelligence Advanced Research Projects Activity (IARPA) via contract FA8650-19-C-9101, as well as the gift fund from NGK Spark Plugs. The author would also like to thank Nanova Environmental, Inc. for providing the chromatograms.

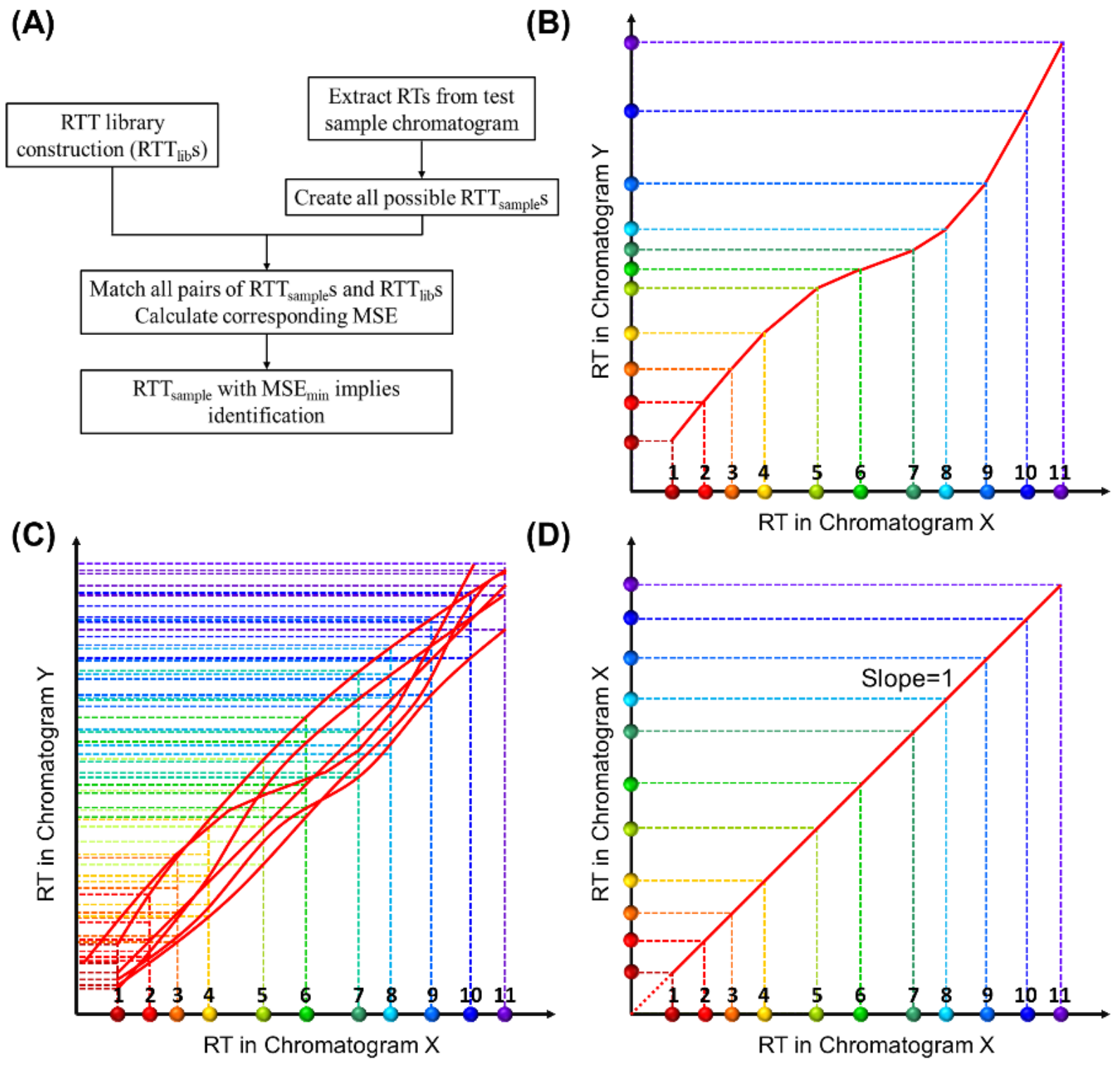

**Figure 1.** Conceptual illustration of retention time trajectory (RTT) and corresponding library construction. **(A)** Algorithm flow chart **(B)** RTT of a chromatogram (Chromatogram Y) - $RTT_Y$. The X-axis represents the retention time ($RT_X$) from one of the chromatograms, which we call Chromatogram X. The colored dots along the X-axis represent different compounds and are numerically labelled as 1, 2, 3, … 11. Similarly, the Y-axis represents the retention time, $RT_Y$, in another chromatogram, Chromatogram Y. The entire set of coordinates, ($RT_{X,\text{compound } i}$, $RT_{Y,\text{compound } i}$), where "compound *i*" refers to a specific compound, form a trajectory (red curve) in 2D. **(C)** Conceptual illustration of the $RTT_{lib}$ library, which is composed of multiple chromatograms obtained under various experimental conditions (column temperature ramping profiles, and carrier gas flow rate, *etc.*) using a mixture containing all target compounds (and internal standards if needed). **(D)** RTT of Chromatogram X ($RTT_X$), where the retention time values along the Y-axis match exactly those along the X-axis. (*i.e.*, the slope of $RTT_X$ is unity)

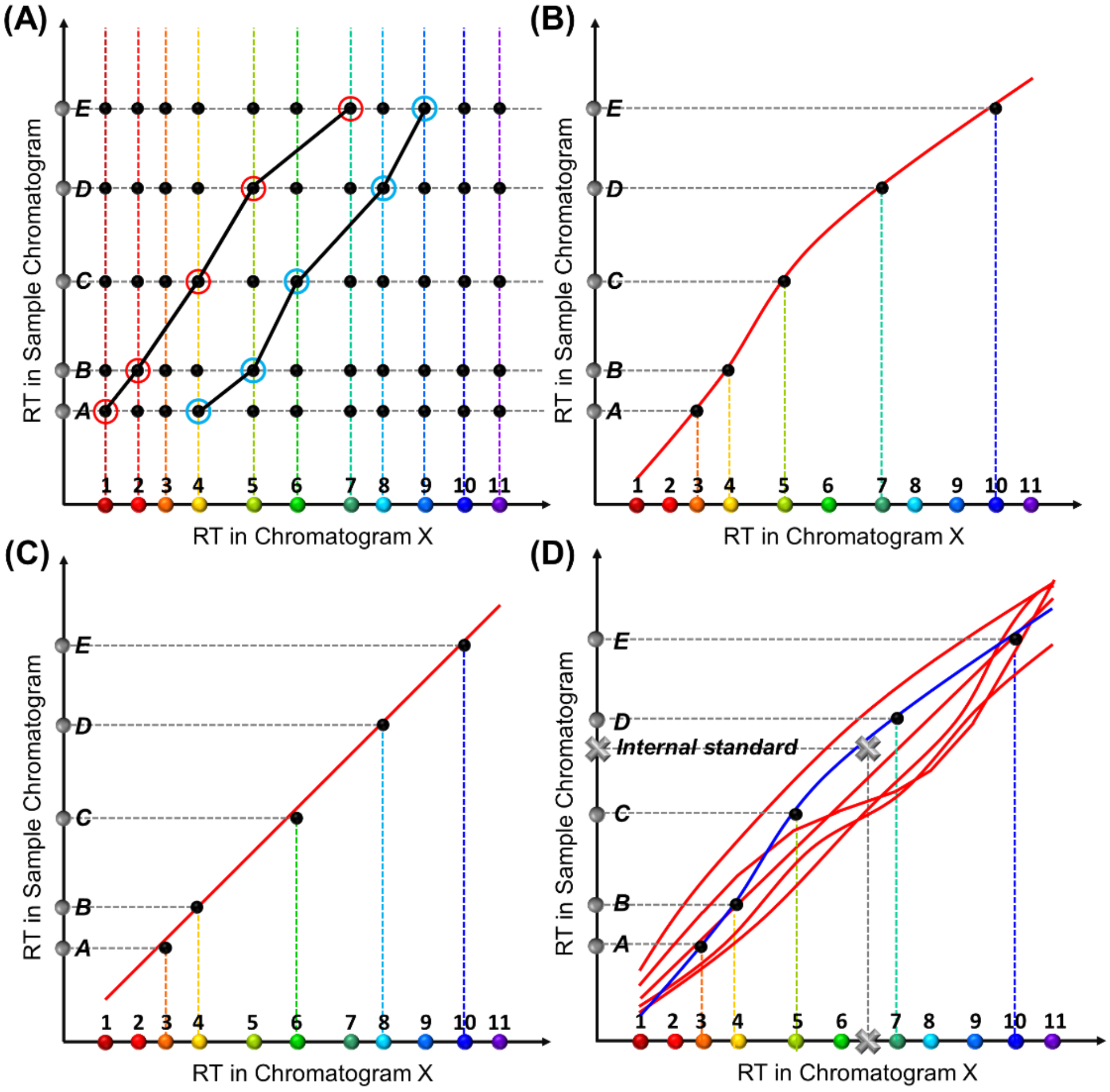

**Figure 2.** Conceptual illustration of generation of an RTT for a test sample, $RTT_{sample}$. **(A)** A retention time matrix formed by $RT_X$ and $RT_{sample}$ of the same compounds. Suppose there are a total of $N_{tgt}$ target analytes (colored dots along the X-axis) and a total of $N_{sample}$ peaks in the sample chromatogram (grey dots denoted from A to E along the Y-axis), then a matrix of $N_{tgt} \times N_{sample}$ coordinates (black dots) can be formed. The two black curves show two exemplary $RTT_{sample}$s formed by connecting one dot on each row. **(B)** and **(C)** Two sets of black dots fall simultaneously on two different $RTT_{lib}$s, suggesting that two $RTT_{sample}$s are found to match two different $RTT_{lib}$s, which leads to two different chemical identification results. Accordingly, Peaks A-E are identified as Compounds 3, 4, 5, 7, and 10 in (B), and Compounds 3, 4, 6, 8 and 10 in (C), respectively. **(D)** An internal standard (grey cross), anchors one of the $RTT_{lib}$s (plotted in blue), thus producing unique chemical identification.

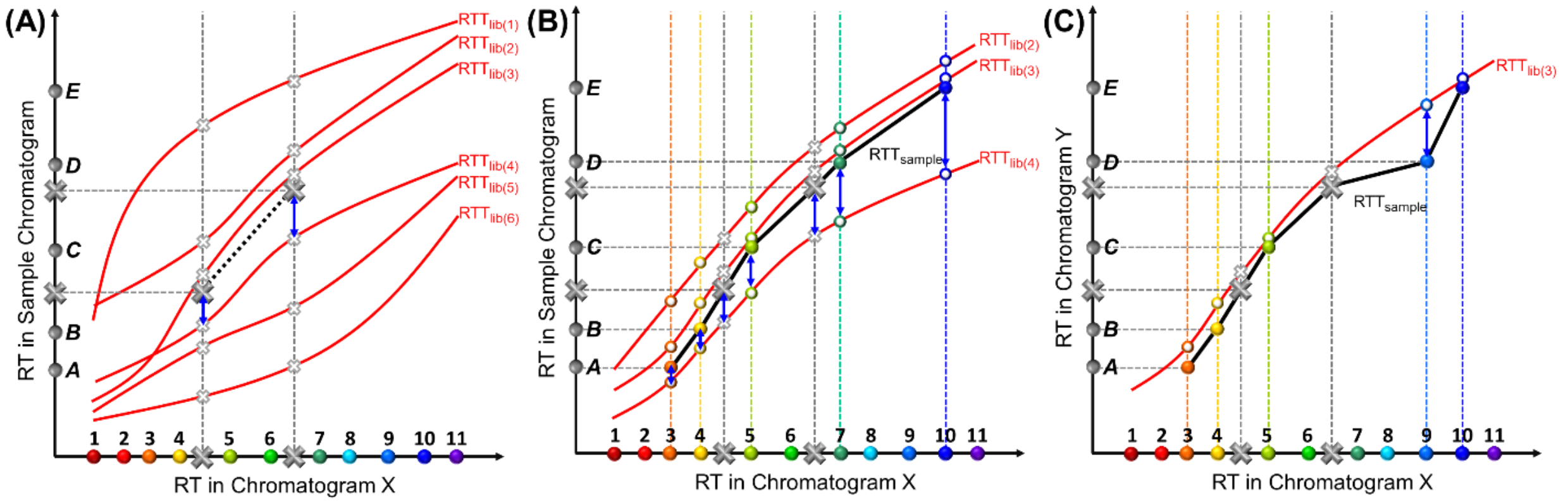

**Figure 3.** *MSR* or *SSR* calculation. **(A)** Illustration of calculating the $SSR^{std}$ between an RTT$_{sample}$ and RTT$_{lib(4)}$ using the internal standards for RTT$_{lib}$s screening. All RTT$_{sample}$s must pass through the two large solid grey crosses, which mark the RTs of the internal standards in the sample chromatogram. The smaller hollow grey crosses mark the RTs of the internal standards in the RTT$_{lib}$s. The internal standard retention time residuals are marked by blue arrows. Half of the RTT$_{lib}$s with the lowest SSR$_{std}$s (*i.e.*, RTT$_{lib(2)}$, RTT$_{lib(3)}$, and RTT$_{lib(4)}$) are kept and used for the next step described in Figure (B). **(B)** Calculation of the *MSR* between an RTT$_{sample}$ and an RTT$_{lib}$ screened in (A) using both target compounds and internal standards. This process quantifies the similarities between the two RTTs under comparison (*i.e.*, one RTT$_{sample}$ and one RTT$_{lib}$). The RTT$_{sample}$ (black curve) is formed by pairing Peaks A to E with target Compounds 3, 4, 5, 7, and 10. The retention time residual of the same compound is marked by a blue arrow. **(C)** Interferent identification. One RTT$_{sample}$ (black curve) is formed by respectively pairing Peaks A to E with target Compounds 3, 4, 5, 9, and 10, and is further compared with RTT$_{lib(3)}$ (red curve). The retention time residual between Peak D and Compound 9 is much larger than those of other pairs. The physical interpretation is that the coordinates associated with Peaks A, B, C, and E in one RTT$_{sample}$ closely match with RTT$_{lib(3)}$. However, Peak D deviates far from RTT$_{lib(3)}$, likely identifying Peak D as an interferent, and identifying Peaks A, B, C, and E as Compounds 3, 4, 5, and 10, respectively.

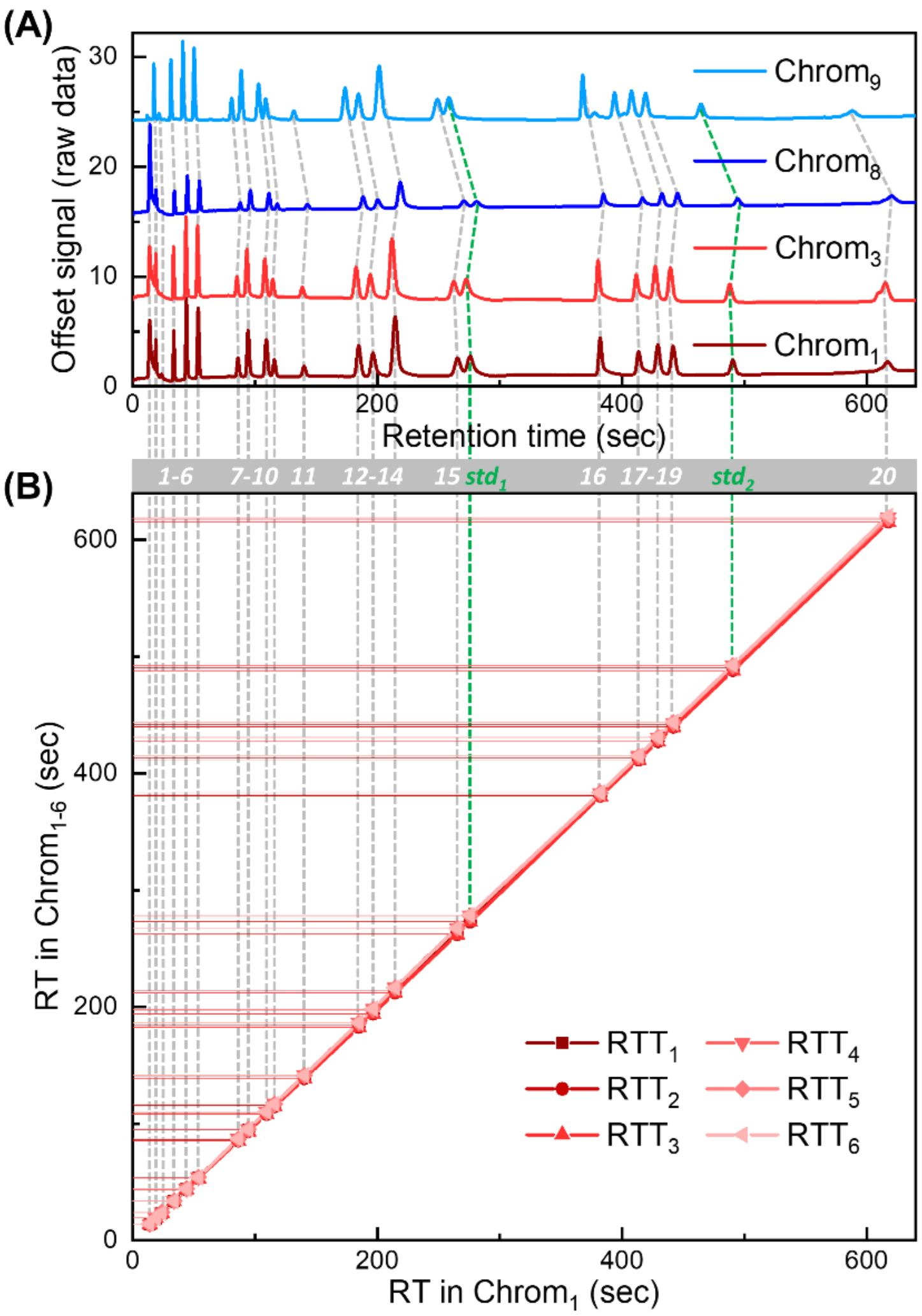

**Figure 4.** Experimentally generated chromatograms and corresponding $RTT_{lib}$s for algorithm validation. **(A)** Four exemplary chromatograms ($Chrom_1$, $Chrom_3$, $Chrom_8$, and $Chrom_9$) obtained experimentally. There are a total of 22 detected peaks in each chromatogram. 2 out of the 22 peaks are treated as internal standards (labelled as $std_1$ and $std_2$ in green). The remaining peaks are treated as target compounds for peak identification in targeted analysis. Peaks of the same compounds are labelled with compound IDs from 1-20 (grey bar). **(B)** $RTT_{lib}$ library formed by $RTT_{lib}$s ($RTT_{1\text{-}6}$) generated from $Chrom_{1\text{-}6}$. The X-axis represents retention time in $Chrom_1$. The Y-axis represents retention times in $Chrom_{1\text{-}6}$. The retention time deviation (ΔRT) of $Chrom_{1\text{-}9}$ against the RT in $Chrom_1$ is plotted in Figure S5.

(A)

Test data generated from Chrom$_7$

**Test 1**

| Retention time (sec) | | | 13.8 | 19 | 33.5 | 43.7 | 85.6 | 108.5 | 115 | 183.6 | 213.3 | 412.7 | 616 |
|---|---|---|---|---|---|---|---|---|---|---|---|---|---|
| **Compound ID** | | | 1 | 2 | 4 | 5 | 7 | 9 | 10 | 12 | 14 | 17 | 20 |
| **Ranking** | **MSR** | **Accuracy** | **Individual peak identification result** | | | | | | | | | | |
| 1st | 0.57 | 100% | 1 | 2 | 4 | 5 | 7 | 9 | 10 | 12 | 14 | 17 | 20 |
| 2nd | 0.71 | 100% | 1 | 2 | 4 | 5 | 7 | 9 | 10 | 12 | 14 | 17 | 20 |
| 3rd | 0.91 | 100% | 1 | 2 | 4 | 5 | 7 | 9 | 10 | 12 | 14 | 17 | 20 |
| 4th | 2.2 | 90.9% | 1 | 3* | 4 | 5 | 7 | 9 | 10 | 12 | 14 | 17 | 20 |

**Test 2**

| Retention time (sec) | | | 19 | 33.5 | 43.7 | 85.6 | 93.8 | 108.5 | 115 | 139.3 | 195.2 | 213.3 | 264 | 381.3 | 412.7 |
|---|---|---|---|---|---|---|---|---|---|---|---|---|---|---|---|
| **Compound ID** | | | 2 | 4 | 5 | 7 | 8 | 9 | 10 | 11 | 13 | 14 | 15 | 16 | 17 |
| **Ranking** | **MSR** | **Accuracy** | **Individual peak identification result** | | | | | | | | | | | | |
| 1st | 0.67 | 100% | 2 | 4 | 5 | 7 | 8 | 9 | 10 | 11 | 13 | 14 | 15 | 16 | 17 |
| 2nd | 0.74 | 100% | 2 | 4 | 5 | 7 | 8 | 9 | 10 | 11 | 13 | 14 | 15 | 16 | 17 |
| 3rd | 1.14 | 100% | 2 | 4 | 5 | 7 | 8 | 9 | 10 | 11 | 13 | 14 | 15 | 16 | 17 |
| 4th | 2.15 | 92.3% | 3* | 4 | 5 | 7 | 8 | 9 | 10 | 11 | 13 | 14 | 15 | 16 | 17 |

**Test 3**

| Retention time (sec) | | | 85.6 | 108.5 | 195.2 | 381.3 | 428.2 |
|---|---|---|---|---|---|---|---|
| **Compound ID** | | | 7 | 9 | 13 | 16 | 18 |
| **Ranking** | **MSR** | **Accuracy** | **Individual peak identification result** | | | | |
| 1st | 0.88 | 100% | 7 | 9 | 13 | 16 | 18 |
| 2nd | 0.94 | 100% | 7 | 9 | 13 | 16 | 18 |
| 3rd | 1.48 | 100% | 7 | 9 | 13 | 16 | 18 |
| 4th | 5.74 | 100% | 7 | 9 | 13 | 16 | 18 |

(B)

Test data generated from Chrom$_8$

**Test 4**

| Retention time (sec) | | | 34.1 | 44.6 | 96.2 | 111.4 | 142.9 | 188.4 | 218.6 | 384.8 | 432.6 |
|---|---|---|---|---|---|---|---|---|---|---|---|
| **Compound ID** | | | 4 | 5 | 8 | 9 | 11 | 12 | 14 | 16 | 18 |
| **Ranking** | **MSR** | **Accuracy** | **Individual peak identification result** | | | | | | | | |
| 1st | 2.42 | 100% | 4 | 5 | 8 | 9 | 11 | 12 | 14 | 16 | 18 |
| 2nd | 2.86 | 100% | 4 | 5 | 8 | 9 | 11 | 12 | 14 | 16 | 18 |
| 3rd | 3.34 | 100% | 4 | 5 | 8 | 9 | 11 | 12 | 14 | 16 | 18 |
| 4th | 5.06 | 88.9% | 4 | 5 | 8 | 10* | 11 | 12 | 14 | 16 | 18 |

**Test 5**

| Retention Time (sec) | | | 87.9 | 111.4 | 218.6 | 384.8 | 432.6 |
|---|---|---|---|---|---|---|---|
| **Compound ID** | | | 7 | 9 | 14 | 16 | 18 |
| **Ranking** | **MSR** | **Accuracy** | **Individual peak identification result** | | | | |
| 1st | 2.99 | 100% | 7 | 9 | 14 | 16 | 18 |
| 2nd | 3.74 | 100% | 7 | 9 | 14 | 16 | 18 |
| 3rd | 4.28 | 100% | 7 | 9 | 14 | 16 | 18 |
| 4th | 7.14 | 80% | 7 | 10* | 14 | 16 | 18 |

**Test 6**

| Retention time (sec) | | | 87.9 |
|---|---|---|---|
| **Compound ID** | | | 7 |
| **Ranking** | **MSR** | **Accuracy** | **Individual peak identification result** |
| 1st | 3.80 | 100% | 7 |
| 2nd | 5.10 | 100% | 7 |
| 3rd | 5.80 | 100% | 7 |
| 4th | 16.75 | 100% | 7 |

**Table 1.** Algorithm experimental design and peak identification results with a sample containing only the target compounds. The peaks listed in Tests 1-3 are generated from $Chrom_7$; the peaks listed in Tests 4-6 are generated from $Chrom_8$. An asterisk "*" denotes peak misidentification.

# Supplementary Information to

# Retention time trajectory matching for target compound peak identification in chromatographic analysis

Wenzhe Zang[†‡], Ruchi Sharma[†‡], Maxwell Wei-Hao Li[†‡§], and Xudong Fan[†‡*]

[†] Department of Biomedical Engineering
University of Michigan,
1101 Beal Avenue, Ann Arbor, Michigan 48109, United States

[‡] Center for Wireless Integrated MicroSensing and Systems (WIMS$^2$)
University of Michigan
Ann Arbor, Michigan 48109, United States

[§] Department of Electrical Engineering and Computer Science
University of Michigan
Ann Arbor, Michigan 48109, United States

*Corresponding author: xsfan@umich.edu

## Section S1. Algorithm validation tests generation

Depending on the number of target compounds, the sample for targeted analysis can be classified as: (1) all target compounds are present, (2) a subset of target compounds are present, and (3) a subset of target compounds as well as interferents are present. In the first scenario (1) the peak identification can be easily achieved as the peaks in the sample chromatogram and in one pre-characterized chromatogram can be paired respectively (and therefore identified) based on the elution order. Peak identifications for samples of the other two scenarios are much more challenging, as the peaks can be paired with any subset of the target compounds. Therefore, in this work, we only discuss and show the details of 11 validation test results that fall under (2) and (3).

In our RTT matching method, the only inputs are the retention times. Therefore, one chromatogram, which is made up of a large 2D array of detection signal intensities vs. time, can be simplified as a list of retention times (Figure S7A). As the peak area and profile do not contribute to the peak identification, multiple algorithm validation tests can simply be generated from an actual chromatogram, which includes all target compounds and internal standards. This can be accomplished by selecting various subsets of the target compounds and adding hypothetical peak positions (*i.e.*, interferents), if needed.

In this work, a total of $3 \times \sum_{i=1}^{20}[\mathrm{C}(20, i)] = 3.15 \times 10^5$ validation tests are generated out of $\mathrm{Chrom}_{7\text{-}9}$, which cover *all* subsets of the 20 target compounds. Peak identifications in all validation tests achieve 100% accuracy. Particularly, 11 validation tests are selected and discussed in details to represent samples of various compositions. Tests 1-3 (Table 1A) are generated from $\mathrm{Chrom}_7$; Tests 4-9 (Tables 1B and S2) are generated from $\mathrm{Chrom}_8$. Tests 10 and 11 (Table S3) are generated from $\mathrm{Chrom}_9$. Tests 5-8 are exemplified in Figures S7B-E, respectively.

## Section S2. Sample chromatogram reconstruction for COW aligning

Correlation optimized warping (COW) algorithm[1,2] is commonly used for chromatogram aligning. In order to compare the peak identification performance of COW aligning with RTT matching, a sample chromatogram (*i.e.*, $Chrom_{sample}$) is reconstructed out of $Chrom_8$, where only peaks listed in the validation test (plus two internal standards) are kept and the rest are replaced with the baseline. Test 5 (a subset of target compounds) and 7 (a subset of target compounds plus one internal standard), which have been validated with the RTT matching approach, are also used to evaluate peak identification with COW aligning.

In this work, the following steps for $Chrom_{sample}$ reconstruction are adopted. First, the open-source adaptive iterative reweighted Penalized Least Squares (airPLS) algorithm[3] is used to correct baseline drifting, and the baseline signal is numerically centered around zero. Second, the signal is smoothed via locally weighted scatterplot smoothing (LOWESS)[4] to further improve the signal-to-noise (S/N) ratio in $Chrom_8$. Next, the chromatogram curve is scanned for peak detection, where peak apex positions, peak heights and endpoints are extracted[5]. Each peak profile can then be fitted by an exponentially modified Gaussian (EMG) equation[6]. Finally, EMG expressions for all target compounds peaks listed in the validation tests (Tests 5 and 7) as well as two internal standards, are added to form a chromatogram ($Chrom_{sample}$) as the input for COW aligning. An artificial EMG expression is adopted to generate the interferent peak in Test 7. The $Chrom_{sample}$s for Tests 5 and 7 are plotted in Figure S8 together with $Chrom_8$.

During COW aligning, $Chrom_1$, where all target compounds and internal standards are present, is treated as the reference chromatogram to be aligned with. Peak identification can be achieved only when the peaks of identical compounds are well aligned between $Chrom_1$ and $Chrom_{sample}$. Multiple COW aligning with various tuning parameters (slack and correlation power) are conducted; the corresponding aligned $Chrom_{sample}$s are plotted in Figures S8B-E and S9B-E. The identification results based on the retention times of the aligned $Chrom_{sample}$ are summarized in Table S4.

## Section S3. Fruit metabolomics datasets

Two publicly available fruit metabolomics datasets from the Metabolights repository (http://www.ebi.ac.uk/metabolights) with identifiers MTBLS99 and MTBLS85 are used to demonstrate the potential of RTT matching for complicated samples. The same datasets were also used previously for fast PTW algorithm validation[7].

The first dataset consists of 23 measurements of a pooled sample that was injected regularly as a quality control (QC) during LC apple extracts measurements. The QC chromatograms are plotted in Figure S10A. The first 7 chromatogram sets (plotted in red) are used for RTT hybridization and subsequent RTT library construction. The remaining 16 chromatograms (plotted in blue) are used to generate validation tests. 40 peaks, which are marked with orange circles, are treated as target compounds. 2 peaks, which are marked with green triangles, are used as internal standards. The remaining peaks show inconsistent elution pattern due to co-elution of different degrees across the chromatograms. They are therefore excluded from the target compound list. Following the same test design in Section S1, we choose target compound subsets containing 40, 30, 20, 10, and 5 target compounds and generate a total of $16 \times [\mathrm{C}(40{,}40) + \mathrm{C}(40{,}30) + \mathrm{C}(40{,}20) + \mathrm{C}(40{,}10) + \mathrm{C}(40{,}5)] = 2.23 \times 10^{12}$ tests. All peaks in the validation tests are correctly identified with the RTT matching approach.

The second datasets are from 14-day LC measurements of carotenoids in grape samples. In the fast PTW work[7], the data was clustered into 14 components according to spectral characteristics. Chromatograms in Component 7, which are plotted in Figure S10B, are used for validation. Similarly, chromatograms in the first 5 days (plotted in red) are used for RTT hybridization and subsequent RTT library construction. The remaining 9 chromatograms (plotted in blue) are used to generate validation tests. 13 peaks (marked with orange circles) are chosen as target compounds, and one peak (marked with green triangle) is used as internal standard. Following the same test design in Section S1, we use all of the target compound subsets (ranging from single compounds to all 13 compounds) for a total of $9 \times \sum_{i=1}^{13}[\mathrm{C}(13, i)] = 73{,}719$ tests. All peaks in the validation tests are correctly identified with the RTT matching approach.

## Section S4. Chromatogram aligning enabled by RTT matching

In the main text, we mainly discussed the concept of RTT matching and its application in peak identification for chromatographic analysis. For visualization purposes, the RTT matching approach is also useful for chromatography profile aligning, which may be assisted by peak reconstruction if needed.

Generally, most chromatogram profile aligning approaches involve selection or generation of a reference chromatogram. All other sample chromatograms are aligned with the reference chromatogram and peaks of the same compound are aligned to the RT in the reference chromatogram. Herein, any $RTT_{lib}$ can be chosen as the reference to extract RTs of target compounds (and internal standards if present). The peaks in the sample chromatogram are first identified with RTT matching and each peak can be shifted to the corresponding RTs in the reference chromatogram.

In order to eliminate the influence of baseline drifting on the peak profile, the sample chromatogram can be reconstructed by fitting each peak with an EMG model after baseline removal, smoothing and peak detection. Methods for each step can be found in Section S2. The aligned sample chromatogram is obtained by summation of all shifted EMG expressions. Illustration of RTT matching based chromatogram aligning is provided in Figure S11, where $Chrom_1$ is used as the reference chromatogram and two sample chromatograms are generated out of $Chrom_8$. $Chrom_{sample1}$ is created through summation and reconstruction all peaks in $Chrom_8$ via EMG, representing a sample with the same analyte composition as the reference chromatogram. Similarly, $Chrom_{sample2}$ is reconstructed only with the peaks listed in Test 5 (Table 1B), representing a sample with only a subset of peaks in the reference chromatogram. $Chrom_{sample1}$ and $Chrom_{sample2}$ are plotted in Figure S11A together with $Chrom_8$, showing that all peak profiles and retention times are well preserved. Aligned chromatograms are shown in Figure S11B and C together with the reference chromatogram (*i.e.*, $Chrom_1$), demonstrating excellent alignment between peaks of the same compounds.

Because peak RTs are the only input for the peak aligning, the RTT matching based chromatogram aligning completely avoids misaligning resulting from disparities in peak size, peak profile, and baseline drift, which are challenging for many other chromatogram aligning approaches to handle.

## Section S5. Glossary of abbreviations and symbols

| Abbreviation or symbols | Description |
|---|---|
| RT | Retention time of a compound, which is equal to the x-coordinate of the apex of the corresponding peak in a chromatogram. |
| ΔRT | The retention time deviation of the same compound in the chromatograms of different runs. |
| RTT | Retention time trajectory, which is made up of discrete points of retention time values of a list of compounds in a measured chromatogram. |
| $\mathrm{RTT_{lib}}$ | Retention time trajectory in a pre-characterized chromatogram. |
| $\mathrm{RTT_{sample}}$ | Retention time trajectory for the chromatogram obtained from a sample under test, which is formed by pairing the detected peaks with a single subset of target compounds. |
| $N_{tgt}$ | Number of target analytes in a pre-determined list. |
| $N_{sample}$ | Number of detected peaks in the chromatogram of the sample under test, with the added internal standards excluded. |
| $N_{inf}$ | Number of interferents in the actual sample under test. |
| $N_{std}$ | Number of internal standard compounds. |
| $n_{lib}$ | Number of pre-characterized RTT ($\mathrm{RTT_{lib}}$) in the library. |
| $n_{sample}$ | Number of all possible RTTs generated from the chromatogram of the actual sample. |
| *SSR* | Sum of squared residuals of RTs from the same compounds between one $\mathrm{RTT_{sample}}$ and one $\mathrm{RTT_{lib}}$. A residual is the difference between the RT (treated as observed value) of a peak in one sample chromatogram and the RT of paired compound in one $\mathrm{RTT_{lib}}$ (treated as predicted value). |
| *MSR* | Mean squared residual, which is *SSR* normalized by the total of number of paired compounds. |

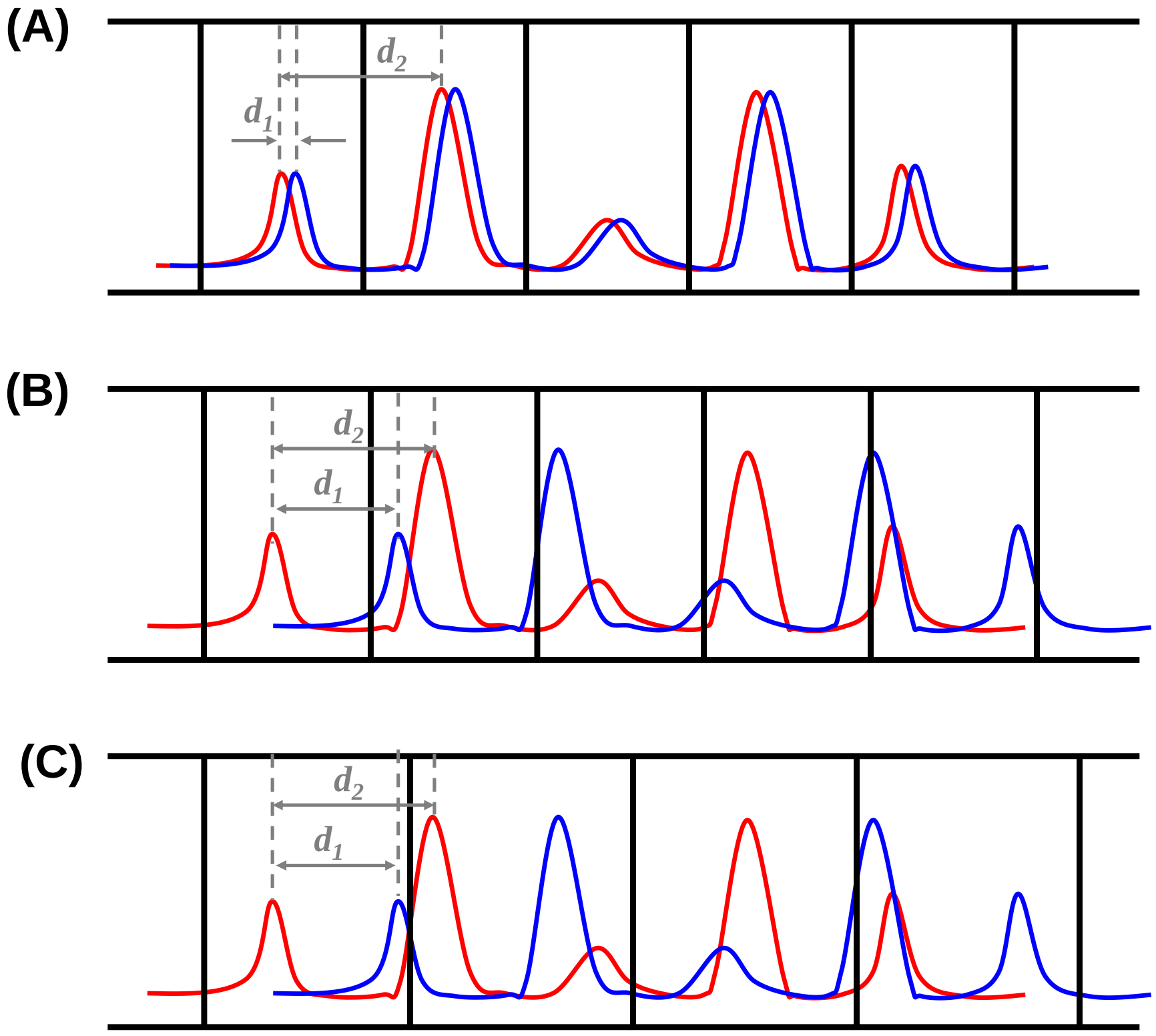

**Figure S1.** Conceptual illustration of the binning approach. **(A)** When drift of the same compound in two different chromatograms (red and blue), $d_1$, is much smaller than the distance between two neighboring peaks in the same chromatogram, $d_2$, peaks of the same compound can be well matched (ending up in the same bin) via appropriate selection of the binning size. However, when $d_1$ and $d_2$ are comparable, as illustrated in **(B)** and **(C)**, peak mismatch occurs regardless of the binning size. **(B)** A narrow binning size incorrectly places the peaks from different compounds (*e.g.*, the first peak of the blue chromatogram and the second peak in the red chromatogram) into the same bin. **(C)** A wide binning size incorporates multiple peaks into the same bin (*e.g.*, the second and the third peak in the red chromatogram), which reduces analysis resolution.

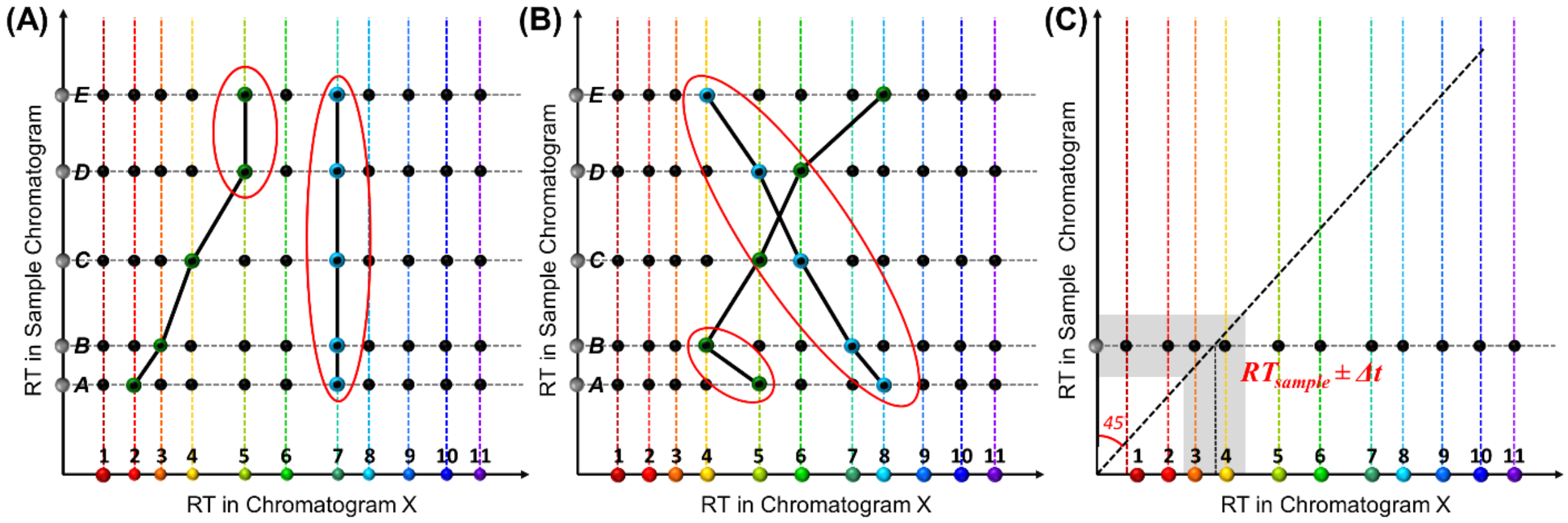

**Figure S2.** Rules to eliminate impossible $RTT_{sample}$s. **(A)** Two examples (circled in red) of impossible $RTT_{sample}$s, in which there is at least one vertical segment between two consecutive coordinates (black dots). A vertical segment means that one target compound is simultaneously assigned to multiple peaks in the chromatogram obtained from the sample under test. **(B)** Two examples (circled in red) of impossible $RTT_{sample}$s with an incorrect elution orders, in which at least one segment between two consecutive coordinates (black dots) has a negative slope. **(C)** Only the coordinates (black dots) falling in the grey-shaded area need to be considered in order to expedite computation.

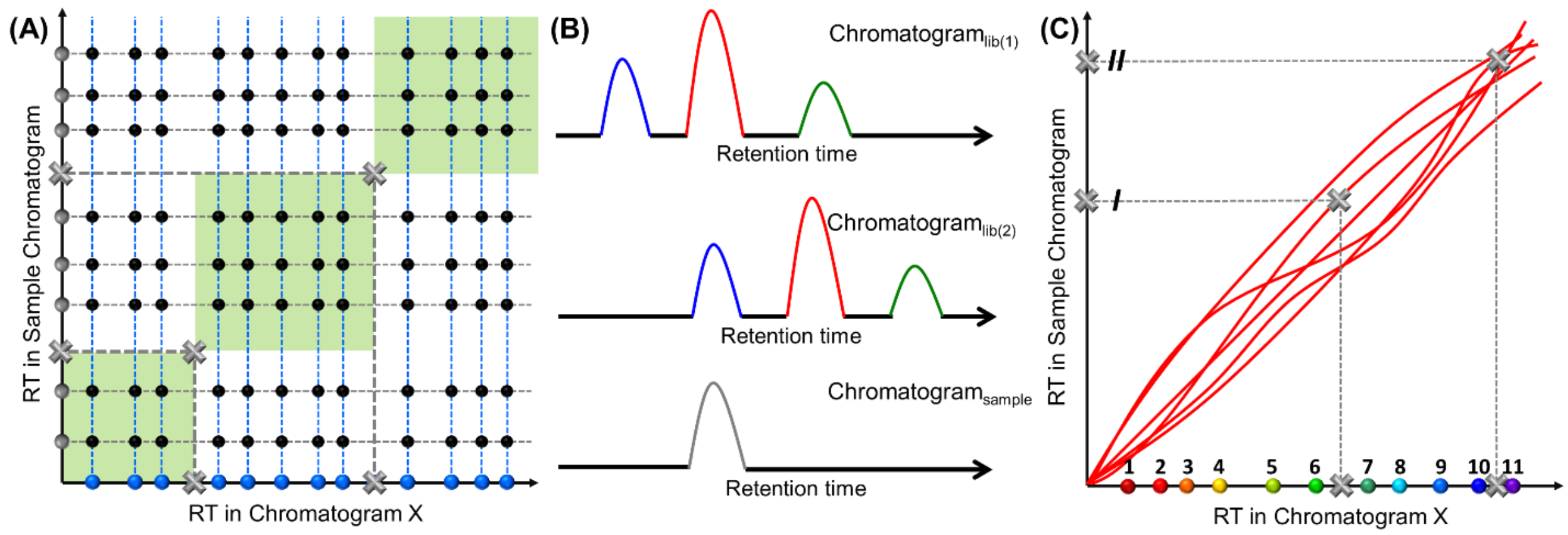

**Figure S3. (A)** Internal standards (grey crosses) divide the 2D diagram into multiple sub-sections (green regions). All possible RTT$_{sample}$s must go through the grey crosses. Therefore, only the black dots (coordinates) falling within these regions can be used to form possible RTT$_{sample}$ candidates. **(B)** RT-based identification of a sample containing a single species is impossible, since the same RT value can also result from drifting neighboring peaks. **(C)** Internal standards should be strategically positioned in the regions where variations in RTTs are more drastic. For example, Compound *I* is more effective than Compound *II* as an anchor for the trajectories because RTTs vary more significantly in the Compound *I* region than in the Compound *II* region.

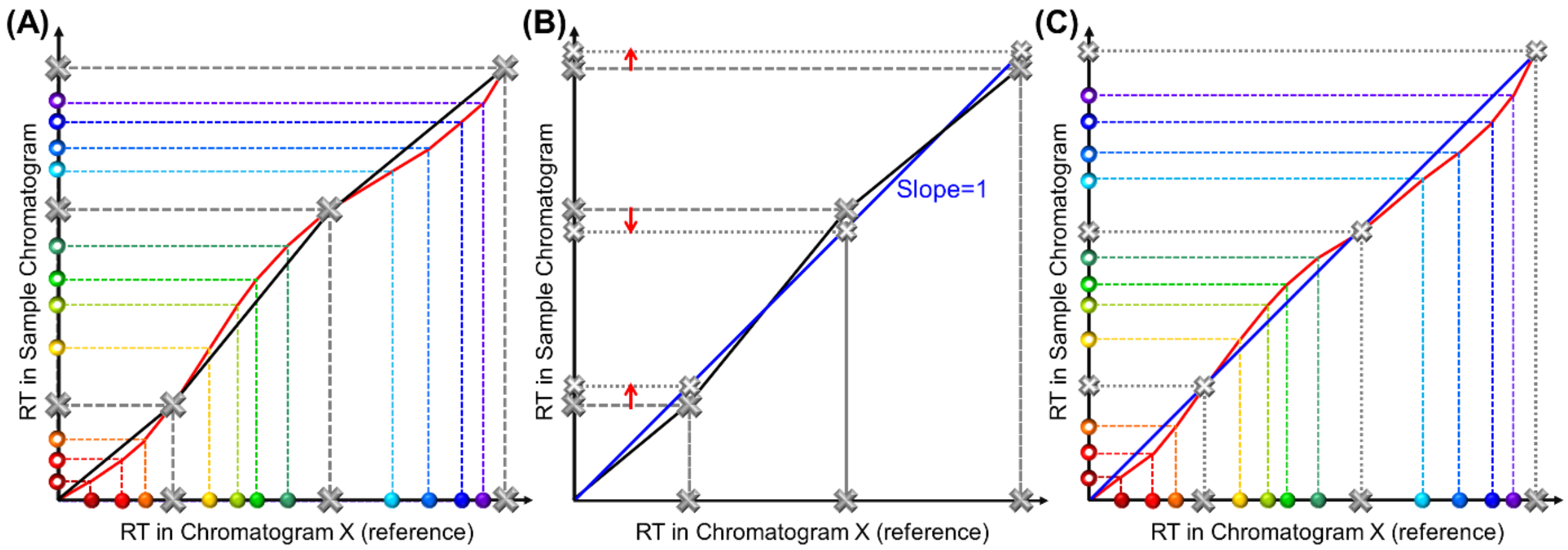

**Figure S4.** Conceptual illustration of peak aligning with the internal standard based linear stretching/compressing approach using the RTT 2D diagram. **(A)** Retention times in the sample chromatogram and those in Chromatogram X (*i.e.*, "reference chromatogram" used in the linear stretching/compressing approach) form the $RTT_{sample}$ shown as the red curve. The first step of linear stretching/compressing is to connect the internal standards (solid gray crosses) using linear lines (marked as black straight lines). **(B)** The second step is to change the slope of those black lines to unity in order to match the retention times in the reference chromatogram (*i.e.*, Chromatogram X). Note that in the RTT 2D diagram, Chromatogram X is represented by a straight line with a slope of one (see the blue line, where the internal standards are marked as small hollow gray crosses). The red arrows point to the warping direction (stretching or compressing). Stretching/compressing is easily interpreted as a slope change (*i.e.*, the slopes in the three segments are all changed to unity) in the RTT 2D diagram. **(C)** A new $RTT_{sample}$ (red curve) is formed from the original $RTT_{sample}$ in (A) after taking into account the slope change in each segment described in (B). Apparently, the new $RTT_{sample}$ still does not overlap with the blue curve (*i.e.*, the reference chromatogram). Consequently, there still exist differences between the retention times in the sample chromatogram and those in the reference chromatogram, which leads to misidentification of the peaks in the sample chromatogram.

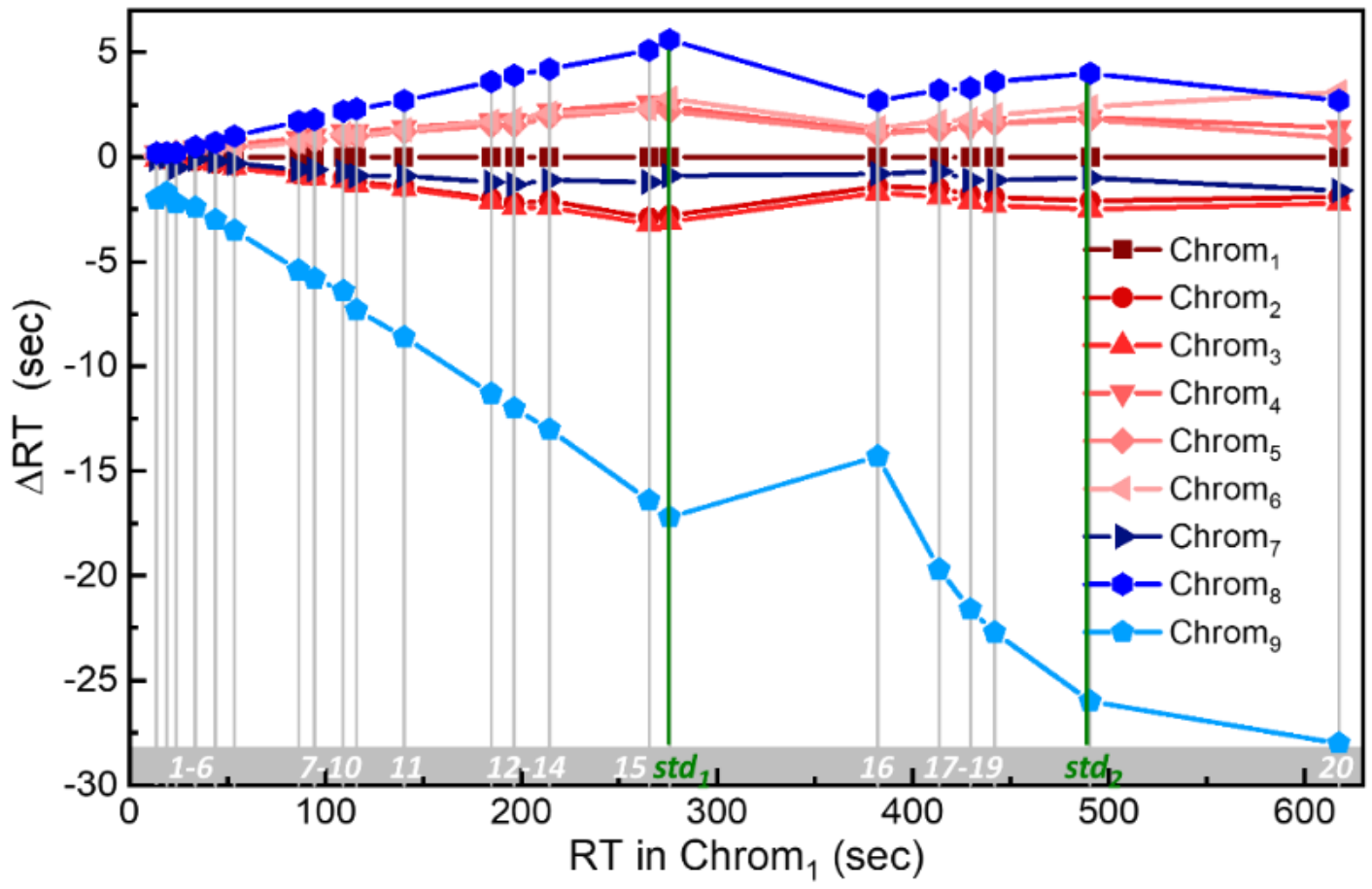

**Figure S5.** Retention time deviation (ΔRT) of $Chrom_{1-9}$ against the RT in $Chrom_1$. The X-axis represents the retention time obtained from $Chrom_1$. ΔRT along the Y-axis is obtained from the RT in $Chrom_{1-9}$ minus the RT of the same compound in $Chrom_1$. A positive (negative) deviation indicates that the corresponding compound elutes later (earlier) than in $Chrom_1$. ΔRT = 0 for all compounds in $Chrom_1$. ΔRT for all chromatograms, except for $Chrom_1$, are highly non-linear.

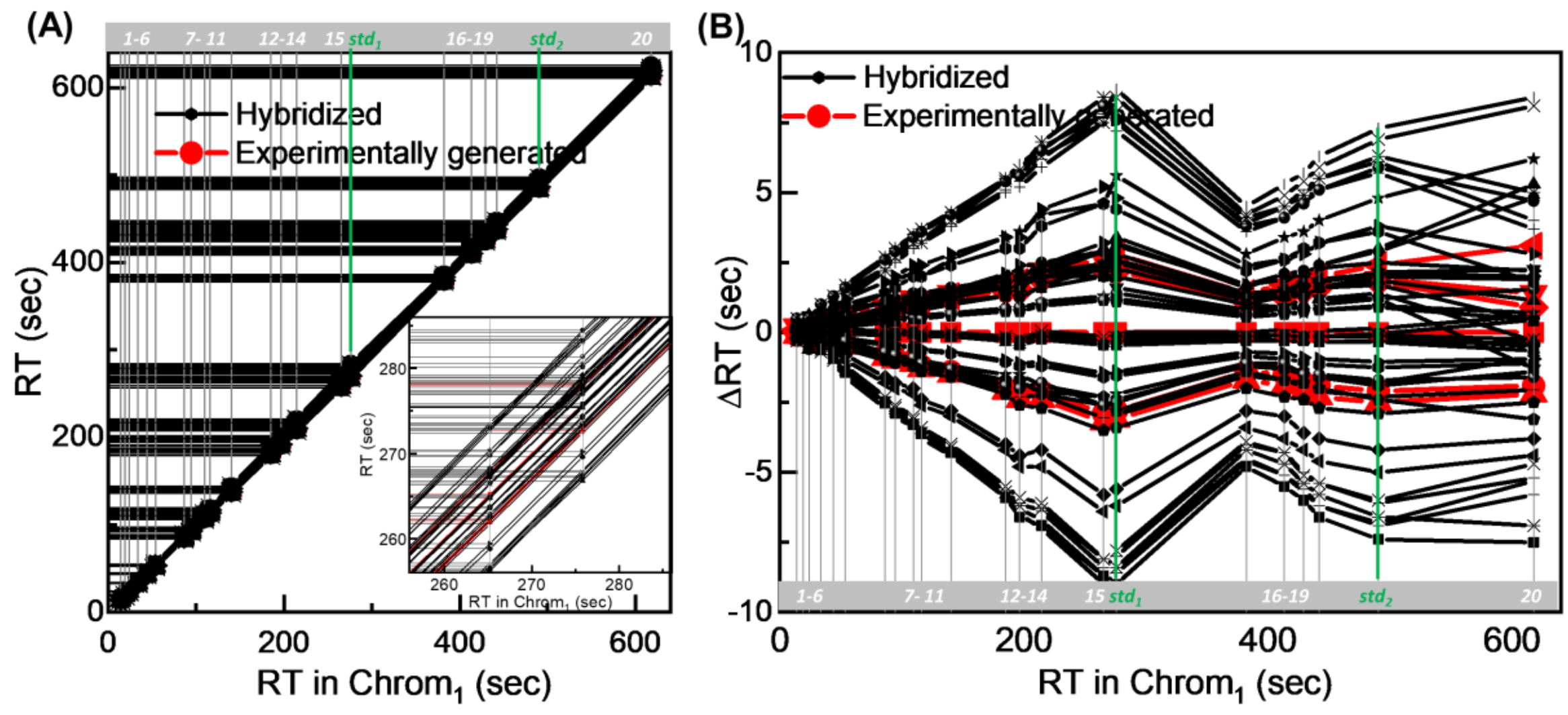

**Figure S6.** Demonstration of $RTT_{lib}$s hybridization. **(A)** RTT library formed by experimentally generated $RTT_{lib}$s ($RTT_{1\text{-}6}$, red) and linearly hybridized $RTT_{lib}$s (black) based on $RTT_{1\text{-}6}$. **(B)** Corresponding retention time deviations (ΔRTs) against RTs in $Chrom_1$.

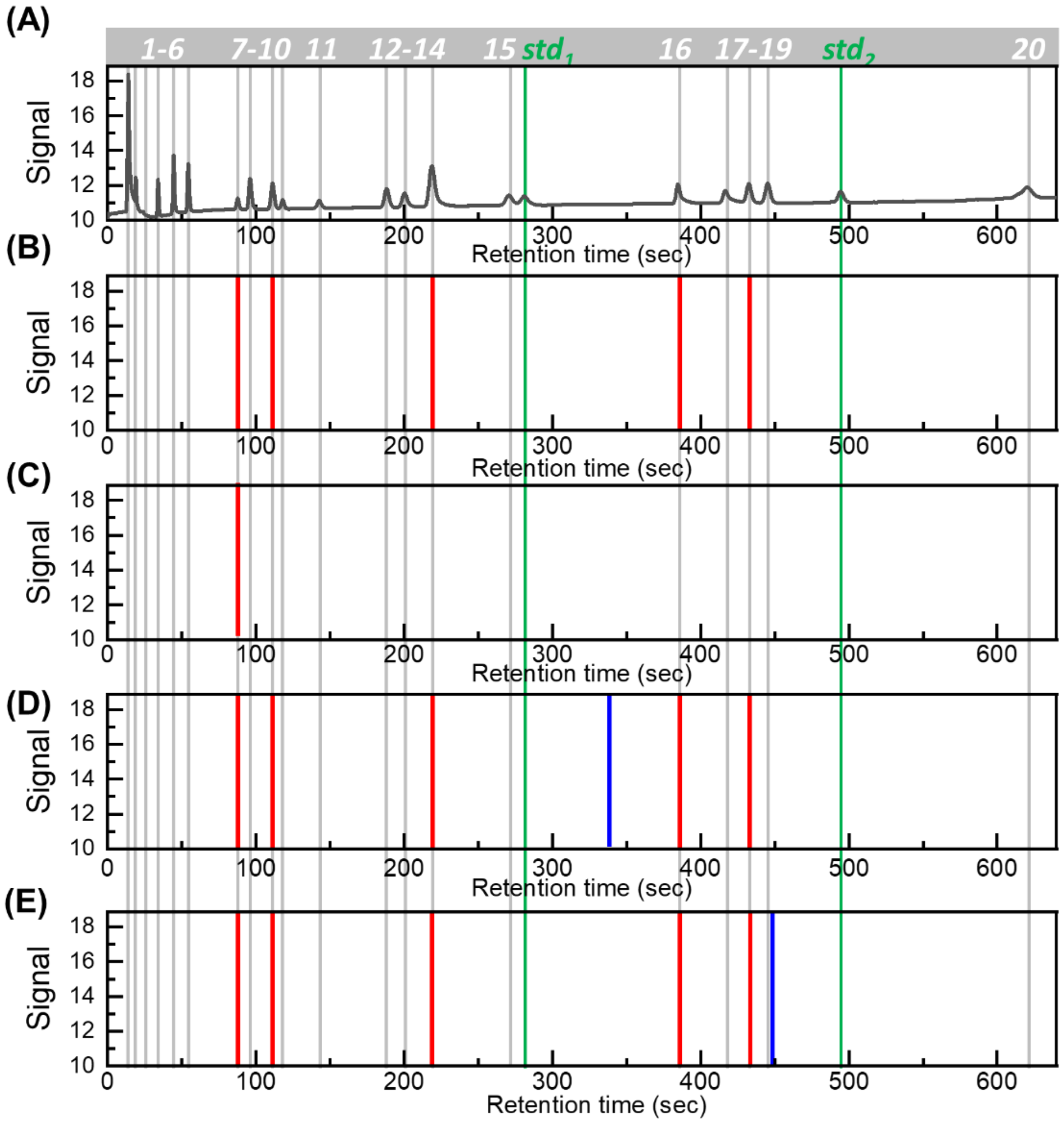

**Figure S7.** Illustration of algorithm validation tests design using $Chrom_8$. **(A)** Experimentally generated $Chrom_8$, where all target compounds and internal standards are present. Peak positions are marked with grey lines. Corresponding peak IDs are listed in the grey bar above. **(B-E)** Illustration of Tests 5-8 generated from $Chrom_8$. The red lines mark the peaks involved in the validation tests. The green lines mark the positions of the two internal standards. The peaks marked in grey are eliminated. Artificially added interferents are marked in blue. All red and grey lines are aligned with the peak apexes in $Chrom_8$ in (A).

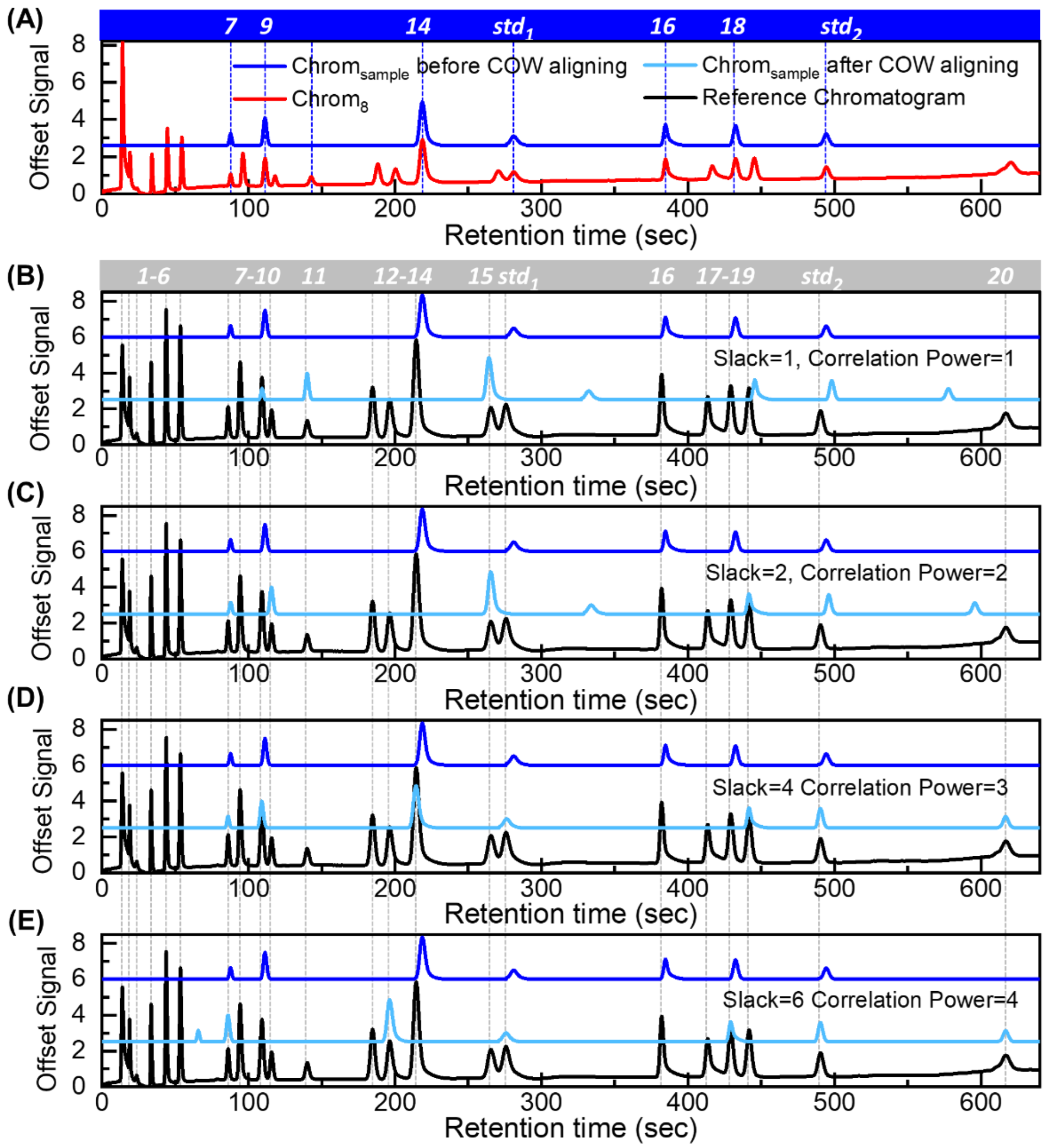

**Figure S8.** Peak identification using COW aligning with a sample containing a subset of target compounds (Test 5). **(A)** $Chrom_8$ and $Chrom_{sample}$ reconstructed from $Chrom_8$ with the peaks in Test 5. Peak retention times before aligning are listed in Table S2. Both peak positions and peak profiles are well preserved in $Chrom_{sample}$ after reconstruction. The corresponding peak IDs are marked in the top blue bar. $Chrom_{sample}$ is vertically shifted for clarity. **(B-E)** Reference chromatogram (*i.e.*, $Chrom_1$), unaligned $Chrom_{sample}$, and aligned $Chrom_{sample}$ using various COW tuning parameters (slack and correlation power). Peak positions and corresponding peak IDs are labelled in the top grey bar. Multiple peaks are aligned to incorrect peaks in the reference chromatogram, leading to misidentification. Identification results are summarized in Table S4A. Unaligned $Chrom_{sample}$ and aligned $Chrom_{sample}$ are vertically shifted for clarity.

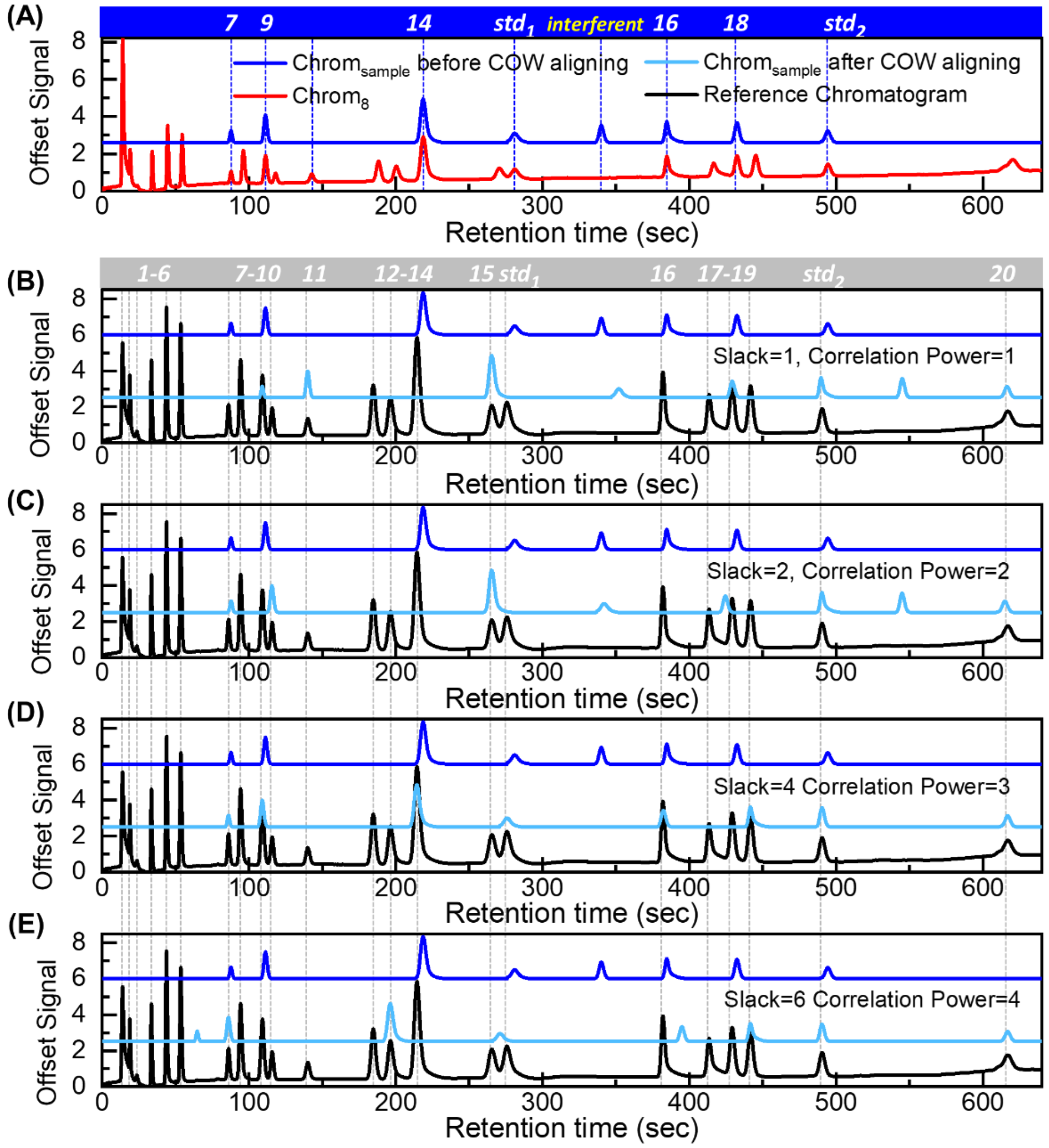

**Figure S9.** Peak identification using COW aligning with a sample containing a subset of target compounds plus one interferent (Test 7). **(A)** $Chrom_8$ and $Chrom_{sample}$ reconstructed from $Chrom_8$ with the target peaks in Test 7. One interferent peak is artificially added at 340 sec with its peak profile generated by an EMG. Peak retention times before aligning are listed in Table S2. Both peak positions and peak profile are well preserved for the target compounds in $Chrom_{sample}$ after the reconstruction. Corresponding peak IDs are marked in the blue bar above. $Chrom_{sample}$ is vertically shifted for clarity. **(B-E)** The reference chromatogram (*i.e.*, $Chrom_1$), unaligned $Chrom_{sample}$, and aligned $Chrom_{sample}$ using various COW tuning parameters (slack and correlation power). Peak positions and corresponding peak IDs are labelled in the top grey bar ~~above~~. Multiple peaks are aligned to the incorrect peaks in the reference chromatogram, leading to misidentification. Identification results are summarized in Table S4B. Unaligned $Chrom_{sample}$ and aligned $Chrom_{sample}$ are vertically shifted for clarity.

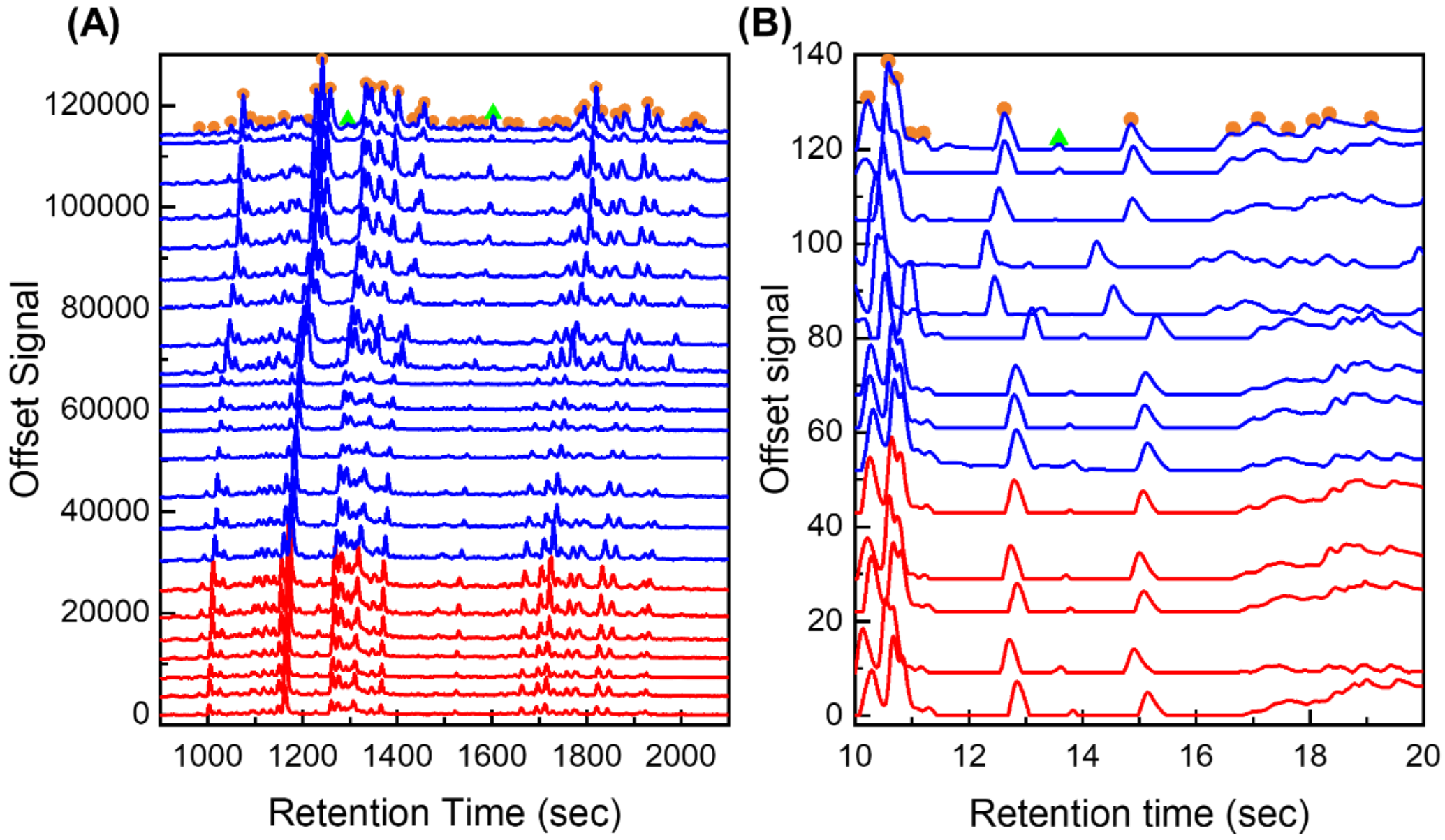

**Figure S10.** Fruit metabolomics chromatograms for RTT peak identification verification. Chromatograms plotted in red are used for RTT library construction and blue ones are used for verification tests design. The orange dots mark the peaks that are treated as target compounds. The green triangles mark the peaks used as internal standards. **(A)** A pooled sample used as QC during the apple extracts measurement. **(B)** Component 7 of carotenoids in grape samples. A detailed sample description can be found in Ref. 7.

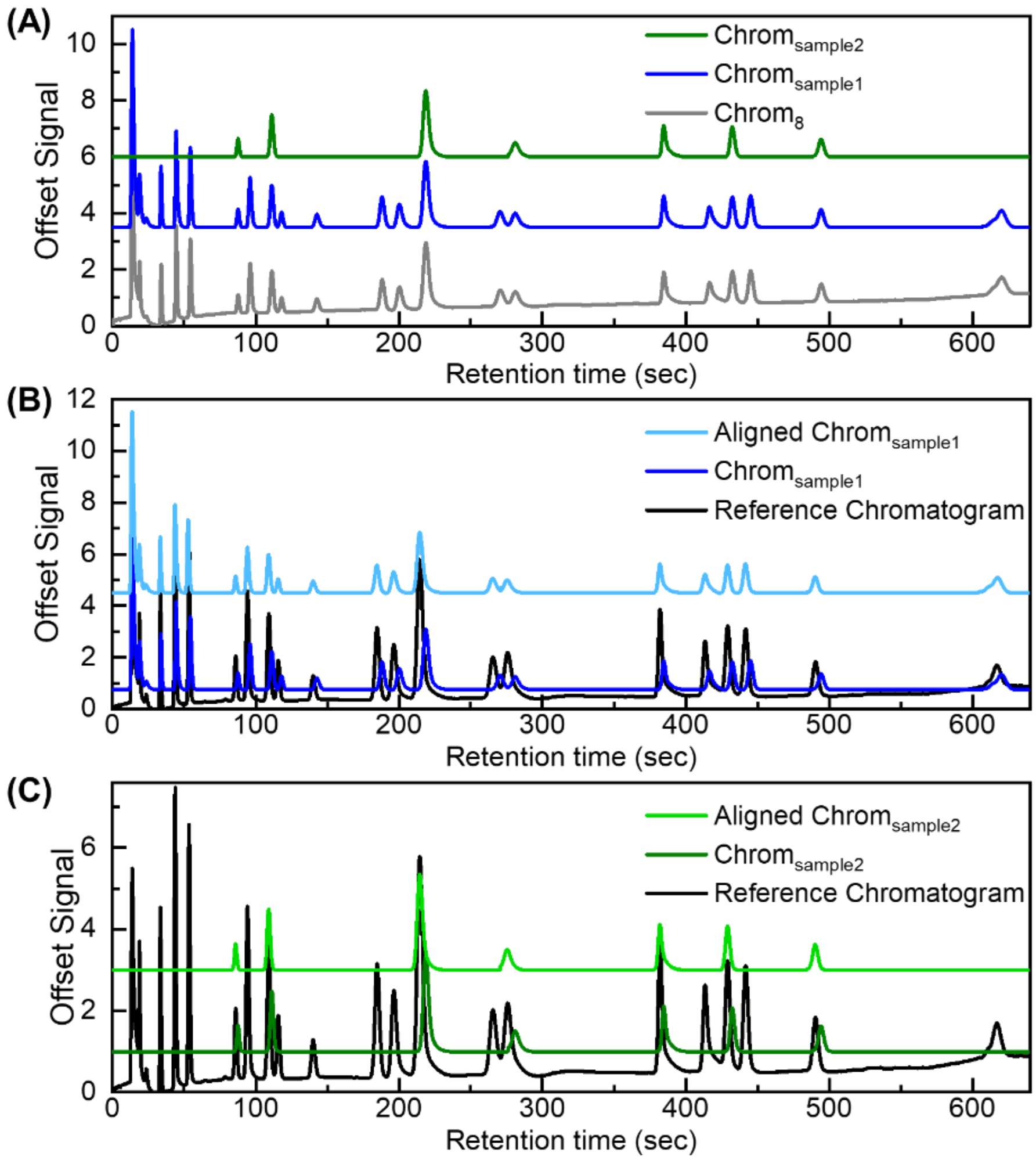

**Figure S11.** Chromatogram aligning enabled by RTT matching. **(A)** Two sample chromatograms (Chrom$_{sample1}$ and Chrom$_{sample2}$) generated from Chrom$_8$. Chrom$_{sample1}$ is reconstructed by fitting all peaks in Chrom$_8$ with EMGs, whereas Chrom$_{sample2}$ only keeps the peaks listed in Test 5. Both retention times and peak profiles are well preserved for all target compounds in Chrom$_{sample}$ after reconstruction. **(B)** and **(C)** Sample chromatograms (Chrom$_{sample1}$ and Chrom$_{sample2}$) before and after the alignment. Chrom$_1$ is used as the reference chromatogram to extract peak RTs for alignment.

| Retention Time (sec) | Compound ID | Compound Name |
|---|---|---|
| 13.9 | 1 | Unknown |
| 19 | 2 | 1,1-Dichloroethene |
| 23.7 | 3 | Unknown |
| 33.6 | 4 | *cis*-1,2-Dichloroethene |
| 43.9 | 5 | Benzene |
| 53.6 | 6 | Trichloroethylene |
| 86.2 | 7 | *cis*-1,3-Dichloropropene |
| 94.4 | 8 | Toluene |
| 109.2 | 9 | Tetrachloroethylene |
| 115.9 | 10 | *trans*-1,3-Dichloropropene |
| 140.2 | 11 | 1,2-Dibromoethane |
| 184.8 | 12 | Chlorobenzene |
| 196.5 | 13 | Ethylbenzene |
| 214.4 | 14 | *m,p*-Xylene |
| 265.2 | 15 | *o*-Xylene |
| 275.8 | $std_1$ | Styrene |
| 382.1 | 16 | 1,3,5-Trimethylbenzene |
| 413.4 | 17 | 1,2,4-Trimethylbenzene |
| 429.3 | 18 | 1,3-Dichlorobenzene |
| 441.8 | 19 | 1,4-Dichlorobenzene |
| 490.5 | $std_2$ | 1,2-Dichlorobenzene |
| 617.6 | 20 | Hexachloro-1,3-Butadiene |

**Table S1.** Peak retention times, assigned compound IDs, and compound names in $Chrom_1$. The same elution order holds for all chromatograms discussed in this work.

**Test data generated from $Chrom_8$**

**Test 7**

| **Retention time (sec)** | | | 87.9 | 111.4 | 218.6 | 340 | 384.8 | 432.6 |
|---|---|---|---|---|---|---|---|---|
| **Compound ID** | | | 7 | 9 | 14 | Interferent | 16 | 18 |
| **Ranking** | **MSR** | **Accuracy** | **Individual peak identification result** | | | | | |
| 1st | 2.99 | 100% | 7 | 9 | 14 | Interferent | 16 | 18 |
| 2nd | 3.74 | 100% | 7 | 9 | 14 | Interferent | 16 | 18 |
| 3rd | 4.28 | 100% | 7 | 9 | 14 | Interferent | 16 | 18 |
| 4th | 7.14 | 83.3% | 7 | 10* | 14 | Interferent | 16 | 18 |

**Test 8**

| **Retention time (sec)** | | | 87.9 | 111.4 | 218.6 | 384.8 | 432.6 | 449 |
|---|---|---|---|---|---|---|---|---|
| **Compound ID** | | | 7 | 9 | 14 | 16 | 18 | Interferent |
| **Ranking** | **MSR** | **Accuracy** | **Individual peak identification result** | | | | | |
| 1st | 2.99 | 100% | 7 | 9 | 14 | 16 | 18 | Interferent |
| 2nd | 3.74 | 100% | 7 | 9 | 14 | 16 | 18 | Interferent |
| 3rd | 4.11 | 83.3% | 7 | 9 | 14 | 16 | 18 | 19* |
| 4th | 4.28 | 100% | 7 | 9 | 14 | 16 | 18 | Interferent |

**Test 9**

| **Retention time (sec)** | | | 34.1 | 44.6 | 62 | 96.2 | 111.4 | 142.9 | 188.4 | 218.6 | 384.8 | 395 | 432.6 |
|---|---|---|---|---|---|---|---|---|---|---|---|---|---|
| **Compound ID** | | | 4 | 5 | interferent | 8 | 9 | 11 | 12 | 14 | 16 | interferent | 18 |
| **Ranking** | **MSR** | **Accuracy** | **Individual peak identification result** | | | | | | | | | | |
| 1st | 2.42 | 100% | 4 | 5 | interferent | 8 | 9 | 11 | 12 | 14 | 16 | interferent | 18 |
| 2nd | 2.86 | 100% | 4 | 5 | interferent | 8 | 9 | 11 | 12 | 14 | 16 | interferent | 18 |
| 3rd | 3.34 | 100% | 4 | 5 | interferent | 8 | 9 | 11 | 12 | 14 | 16 | interferent | 18 |
| 4th | 5.06 | 90.9% | 4 | 5 | interferent | 8 | 10* | 11 | 12 | 14 | 16 | interferent | 18 |

**Table S2.** Algorithm validation tests and the corresponding peak identification results. The sample consists of both target compounds and interferents. Retention times for target compounds are generated from $Chrom_8$. An interferent at 340 s in Test 7 and at 449 s in Test 8 are added artificially. In Test 9, two interferents are artificially added at 62 s and 385 s. An asterisk "*" denotes peak misidentification.

**Test data generated from $Chrom_9$**

**Test 10**

| **Retention time (sec)** | | | 11.92 | 21.52 | 31.2 | 40.8 | 50 | 80.8 | 108.8 | 131.6 | 173.6 | 248.8 | 393.6 |
|---|---|---|---|---|---|---|---|---|---|---|---|---|---|
| **Compound ID** | | | 1 | 3 | 4 | 5 | 6 | 7 | 10 | 11 | 12 | 15 | 17 |
| **Ranking** | **MSR** | **Accuracy** | **Individual peak identification result (with experimentally generated $RTT_{lib}$s only)** | | | | | | | | | | |
| 1st | 98.99 | 81.8% | 1 | 3 | 4 | 5 | 6 | 7 | 9* | 11 | 12 | 15 | 16* |
| 2nd | 99.13 | 72.7% | 1 | 2* | 4 | 5 | 6 | 7 | 9* | 11 | 12 | 15 | 16* |
| 3rd | 101.48 | 81.8% | 1 | 3 | 4 | 5 | 6 | 7 | 9* | 11 | 12 | 15 | 16* |
| 4th | 101.66 | 72.7% | 1 | 2* | 4 | 5 | 6 | 7 | 9* | 11 | 12 | 15 | 16* |
| **Ranking** | **MSR** | **Accuracy** | **Individual peak identification result (with both experimentally generated and hybridized $RTT_{lib}$s)** | | | | | | | | | | |
| 1st | 1.13 | 100% | 1 | 3 | 4 | 5 | 6 | 7 | 10 | 11 | 12 | 15 | 17 |
| 2nd | 2.55 | 90.9% | 1 | 2* | 4 | 5 | 6 | 7 | 10 | 11 | 12 | 15 | 17 |
| 3rd | 3.16 | 90.9% | 1 | 3 | 4 | 5 | 6 | 7 | 9* | 11 | 12 | 15 | 17 |
| 4th | 3.29 | 90.9% | 2* | 3 | 4 | 5 | 6 | 7 | 10 | 11 | 12 | 15 | 17 |

**Test 11**

| **Retention time (sec)** | | | 31.2 | 50.1 | 88.6 | 108.6 | 173.5 | 248.8 | 419.1 |
|---|---|---|---|---|---|---|---|---|---|
| **Compound ID** | | | 4 | 6 | 8 | 10 | 12 | 15 | 19 |
| **Ranking** | **MSR** | **Accuracy** | **Individual peak identification result (with experimentally generated $RTT_{lib}$s only)** | | | | | | |
| 1st | 121.41 | 57.1% | 4 | 6 | 7* | 9* | 12 | 15 | 17* |
| 2nd | 122.28 | 57.1% | 4 | 6 | 7* | 9* | 12 | 15 | 18* |
| 3rd | 122.76 | 71.4% | 4 | 6 | 8 | 9* | 12 | 15 | 17* |
| 4th | 123.63 | 71.4% | 4 | 6 | 8 | 9* | 12 | 15 | 18* |
| **Ranking** | **MSR** | **Accuracy** | **Individual peak identification result (with both experimentally generated and hybridized $RTT_{lib}$s)** | | | | | | |
| 1st | 1.64 | 100% | 4 | 6 | 8 | 10 | 12 | 15 | 19 |
| 2nd | 4.58 | 85.7% | 4 | 6 | 8 | 9* | 12 | 15 | 19 |
| 3rd | 7.38 | 85.7% | 4 | 6 | 7* | 10 | 12 | 15 | 19 |
| 4th | 7.40 | 100% | 4 | 6 | 8 | 10 | 12 | 15 | 19 |

**Table S3.** Algorithm validation tests and the corresponding peak identification results when a sample chromatogram has serious drift issues. Retention times for compounds in Test 9 and Test 10 are generated from $Chrom_9$, which drift much more seriously than $Chrom_{7\text{-}8}$ (see Figure S5). In each test, severe peak misidentification occurs when only the experimentally generated $RTT_{lib}$s (*i.e.*, $Chrom_{1\text{-}6}$) are used. In contrast, when the $RTT_{lib}$s generated by the hybridization method are added, our approach can identify the peaks with 100% accuracy (at least for the top result with the smallest *MSR*). An asterisk "*" denotes peak misidentification.

**(A)**

| **RT in *Test 5* (sec)** | 87.9 | 111.4 | 218.6 | 281.4 | 384.8 | 432.6 | 494.5 |
|---|---|---|---|---|---|---|---|
| **Compound ID** | 7 | 9 | 14 | std$_1$ | 16 | 18 | std$_2$ |
| **Individual peak identification result w/ COW aligning (slack=1, correlation power=1), Accuracy=0** | | | | | | | |
| **RT after aligning (sec)** | 109.2 | 140.2 | 265.2 | 449.5 | 497.1 | 497.1 | 575.7 |
| **Peak identification** | 9* | 11* | 15* | Interferent* | Interferent* | Interferent* | Interferent* |
| **Individual peak identification result w/ COW aligning (slack=2, correlation power=2), Accuracy=0** | | | | | | | |
| **RT after aligning (sec)** | 88.2 | 115.9 | 265.2 | 345.8 | 441.8 | 499.2 | 597.3 |
| **Peak identification** | Interferent* | 10 | 15 | Interferent* | 19* | Interferent* | Interferent* |
| **Individual peak identification w/ COW aligning (slack=4, correlation power=3), Accuracy=57.1%** | | | | | | | |
| **RT after aligning (sec)** | 86.2 | 109.2 | 214.4 | 281.4 | 441.8 | 490.5 | 617.6 |
| **Peak identification** | 7 | 9 | 14 | std$_1$ | 19* | std$_2$* | 20* |
| **Individual peak identification result w/ COW aligning (slack=6, correlation power=4), Accuracy=28.6%** | | | | | | | |
| **RT after aligning (sec)** | 65.1 | 86.2 | 196.5 | 281.4 | 429.3 | 490.5 | 617.6 |
| **Peak identification** | Interferent* | 7* | 13* | std$_1$ | 18 | std$_2$* | 20* |

**(B)**

| **RT in *Test 7* (sec)** | 87.9 | 111.4 | 218.6 | 281.4 | 340 | 384.8 | 432.6 | 494.5 |
|---|---|---|---|---|---|---|---|---|
| **Compound ID** | 7 | 9 | 14 | std$_1$ | Interferent | 16 | 18 | std$_2$ |
| **Individual peak identification result w/ COW aligning (slack=1, correlation power=1), Accuracy=0** | | | | | | | | |
| **RT after aligning (sec)** | 109.2 | 140.2 | 265.2 | 354.1 | 429.3 | 490.5 | 545.4 | 617.6 |
| **Peak identification** | 9* | 11* | 15* | Interferent* | 18* | std$_2$ | Interferent* | 20 |
| **Individual peak identification result w/COW aligning (slack=2, correlation power=2), Accuracy=0** | | | | | | | | |
| **RT after aligning (sec)** | 88.5 | 115.9 | 265.2 | 341.9 | 425.1 | 490.5 | 545 | 617.6 |
| **Peak identification** | Interferent* | 10* | 15* | Interferent* | Interferent | std$_2$* | Interferent* | 20* |
| **Individual peak identification result w/ COW aligning (slack=4, correlation power=3), Accuracy=50%** | | | | | | | | |
| **RT after aligning (sec)** | 86.2 | 109.2 | 214.4 | 275.8 | 382.1 | 441.8 | 490.5 | 617.6 |
| **Peak identification** | 7 | 9 | 14 | std$_1$ | 16* | 19* | std$_2$* | 20* |
| **Individual peak identification result w/ COW aligning (slack=6, correlation power=4), Accuracy=12.5%** | | | | | | | | |
| **RT after aligning (sec)** | 64.7 | 86.2 | 196.5 | 271.8 | 394.7 | 441.8 | 490.5 | 617.6 |
| **Peak identification** | Interferent* | 7* | 13* | Interferent* | Interferent | 19* | std$_2$* | 20* |

**Table S4.** Peak identification performance comparison with COW using Test 5 in Table **(A)** and Test 7 in Table **(B)**. Chrom$_1$ is treated as the reference chromatogram to be aligned to. The two sample chromatograms (*i.e.*, Chrom$_{sample}$s), with or without interferent, are generated from Chrom$_8$, as shown in Figures 8A and 9A. RTs and corresponding compound IDs are listed in the first two rows in each table. Peak identifications and RTs after COW with various parameters are summarized in the remaining rows. An asterisk "*" denotes peak misidentification.

| Test 5 | RT before aligning (sec) | 87.9 | 111.4 | 218.6 | 281.4 | 384.8 | 432.6 | 494.5 |
|---|---|---|---|---|---|---|---|---|
| | **Compound ID** | 7 | 9 | 14 | std$_1$ | 16 | 18 | std$_2$ |
| | **Individual peak identification result with linear warping<br>Same internal standards (std$_1$ and std$_2$) are adopted, Accuracy=28.6%** | | | | | | | |
| | **RT after aligning (sec)** | 86.15 | 109.18 | 214.3 | 275.8 | 380.0 | 428.1 | 490.5 |
| | **Peak identification** | Interferent* | Interferent* | Interferent* | std$_1$ | Interferent* | Interferent* | std$_2$ |

| Test 7 | RT before aligning (sec) | 87.9 | 111.4 | 218.6 | 281.4 | 340 | 384.8 | 432.6 | 494.5 |
|---|---|---|---|---|---|---|---|---|---|
| | **Compound ID** | 7 | 9 | 14 | std$_1$ | Interferent | 16 | 18 | std$_2$ |
| | **Individual peak identification result with linear warping<br>Same internal standards (std$_1$ and std$_2$) are adopted, Accuracy=25%** | | | | | | | | |
| | **RT after aligning (sec)** | 86.315 | 109.18 | 214.3 | 275.8 | 334.84 | 380.0 | 428.1 | 490.5 |
| | **Peak identification** | Interferent* | Interferent* | Interferent* | std$_1$ | Interferent* | Interferent* | Interferent* | std$_2$ |

**Table S5.** Peak identification performance comparison with internal standard based linear warping using Tests 5 and 7. Same internal standards (*i.e.*, std$_1$ and std$_2$) are used for both tests. Chrom$_1$ is treated as the reference chromatogram. The sample chromatogram peak list (*i.e.*, Chrom$_{sample}$) is generated from Chrom$_8$. Corresponding compound IDs are listed in the first two rows of each table. Peak identifications and RTs after linear warping are summarized in the remaining rows. An asterisk "*" denotes peak misidentification.

**(A)**

| Compound ID | Reference | | Chrom8 | | | | Chrom9 | | | |
|---|---|---|---|---|---|---|---|---|---|---|
| | RT | Peak height | RT | Peak height | Warped RT, Order=2 | Warped RT, Order=3 | RT | Peak height | Warped RT, Order=2 | Warped RT, Order=3 |
| **1** | 13.9 | 17.65 | 14.1 | 18.37 | 14.14 | 14.42 | 11.9 | 2.36 | 12.47 | 11.26 |
| **2** | 19 | 15.86 | 19.2 | 12.44 | 19.11 | 19.34 | 17.3 | 7.06 | 18.27 | 17.17 |
| **3** | 23.7 | 12.73 | 23.9 | 10.50 | 23.69 | 23.88 | 21.5 | 2.52 | 22.78 | 21.76 |
| **4** | 33.6 | 16.69 | 34.1 | 12.33 | 33.63 | 33.73 | 31.2 | 7.43 | 33.19 | 32.35 |
| **5** | 43.9 | 19.63 | 44.6 | 13.72 | 43.87 | 43.90 | 40.9 | 9.08 | 43.59 | 42.91 |
| **6** | 53.6 | 18.72 | 54.6 | 13.24 | 53.63 | 53.60 | 50.1 | 8.49 | 53.44 | 52.90 |
| **7** | 86.2 | 14.21 | 87.9 | 11.26 | 86.20 | 86.02 | 80.8 | 3.90 | 86.26 | 86.07 |
| **8** | 94.4 | 16.72 | 96.2 | 12.38 | 94.33 | 94.13 | 88.6 | 6.45 | 94.58 | 94.46 |
| **9** | 109.2 | 15.85 | 111.4 | 12.11 | 109.23 | 109.00 | 102.8 | 5.20 | 109.72 | 109.69 |
| **10** | 115.9 | 14.04 | 118.2 | 11.17 | 115.90 | 115.66 | 108.6 | 3.87 | 115.90 | 115.90 |
| **11** | 140.2 | 13.45 | 142.9 | 11.12 | 140.16 | 139.93 | 131.6 | 2.74 | 140.36 | 140.45 |
| **12** | 184.8 | 15.31 | 188.4 | 11.80 | 184.98 | 184.83 | 173.5 | 4.84 | 184.80 | 184.89 |
| **13** | 196.5 | 14.65 | 200.4 | 11.54 | 196.82 | 196.71 | 184.5 | 4.29 | 196.44 | 196.50 |
| **14** | 214.4 | 17.94 | 218.6 | 13.11 | 214.81 | 214.76 | 201.4 | 6.84 | 214.29 | 214.30 |
| **15** | 265.2 | 14.18 | 270.3 | 11.43 | 266.04 | 266.22 | 248.8 | 3.80 | 264.22 | 264.02 |
| **std1** | 275.8 | 14.35 | 281.4 | 11.38 | 277.06 | 277.30 | 258.6 | 3.95 | 274.52 | 274.26 |
| **16** | 382.1 | 16.02 | 384.8 | 12.06 | 380.20 | 380.88 | 367.8 | 6.01 | 388.61 | 387.89 |
| **17** | 413.4 | 14.79 | 416.6 | 11.70 | 412.08 | 412.84 | 393.7 | 4.41 | 415.50 | 414.77 |
| **18** | 429.3 | 15.38 | 432.6 | 12.09 | 428.15 | 428.93 | 407.7 | 4.56 | 430.00 | 429.30 |
| **19** | 441.8 | 15.26 | 445.4 | 12.11 | 441.02 | 441.80 | 419.1 | 4.36 | 441.80 | 441.13 |
| **std2** | 490.5 | 14.00 | 494.5 | 11.64 | 490.50 | 491.20 | 464.5 | 3.36 | 488.66 | 488.31 |
| **20** | 617.6 | 13.86 | 620.3 | 11.89 | 618.09 | 617.60 | 588.1 | 2.76 | 615.21 | 617.60 |

**(B)**

| | | | | | | | | | |
|---|---|---|---|---|---|---|---|---|---|
| Chrom 8 | Test 5 | **CompoundID** | 7 | 9 | 14 | std1 | 16 | 18 | std2 |
| | | **RT** | 87.9 | 111.4 | 218.6 | 281.0 | 384.6 | 432.4 | 494.3 |
| | | **Peak height** | 0.64 | 1.49 | 2.34 | 0.50 | 1.11 | 1.07 | 0.62 |
| | | **Warped RT, Order=2** | 59.10 | 86.20 | 206.67 | 274.11 | 381.95 | 429.96 | 490.50 |
| | | **Warped RT, Order=3** | 239.66 | 109.20 | 28.11 | 196.50 | 429.30 | 386.54 | 68.11 |

| | | | | | | | | | | |
|---|---|---|---|---|---|---|---|---|---|---|
| Chrom 8 | Test 7 | **CompoundID** | 7 | 9 | 14 | std1 | interferent | 16 | 18 | std2 |
| | | **RT** | 87.9 | 111.4 | 218.6 | 281.0 | 340.0 | 384.6 | 432.4 | 494.3 |
| | | **Peak height** | 0.64 | 1.49 | 2.34 | 0.50 | 0.92 | 1.11 | 1.07 | 0.62 |
| | | **Warped RT, Order=2** | 59.04 | 86.20 | 206.86 | 274.33 | 336.36 | 382.10 | 430.03 | 490.41 |
| | | **Warped RT, Order=3** | 121.10 | 86.20 | 23.70 | 52.08 | 115.90 | 184.85 | 275.80 | 415.47 |

| | | | | | | | | | | | | | | | |
|---|---|---|---|---|---|---|---|---|---|---|---|---|---|---|---|
| Chrom 9 | Test 10 | **CompoundID** | 1 | 3 | 4 | 5 | 6 | 7 | 10 | 11 | 12 | 15 | std1 | 17 | std2 |
| | | **RT** | 11.9 | 21.5 | 31.2 | 40.9 | 50.1 | 80.8 | 108.6 | 131.6 | 173.5 | 248.8 | 258.6 | 393.7 | 464.5 |
| | | **Peak height** | 2.36 | 2.52 | 7.43 | 9.08 | 8.49 | 3.90 | 3.87 | 2.74 | 4.84 | 3.80 | 3.95 | 4.41 | 3.36 |
| | | **Warped RT, Order=2** | -0.46 | 10.85 | 22.25 | 33.60 | 44.33 | 79.88 | 111.73 | 137.83 | 184.80 | 267.35 | 277.91 | 419.43 | 490.50 |
| | | **Warped RT, Order=3** | 273.93 | 181.38 | 109.20 | 56.76 | 23.70 | 11.71 | 96.76 | 206.06 | 416.46 | 441.80 | 378.43 | -3416.86 | -8571.08 |

| | | | | | | | | | | | |
|---|---|---|---|---|---|---|---|---|---|---|---|
| Chrom 9 | Test 11 | **CompoundID** | 4 | 6 | 8 | 10 | 12 | 15 | std1 | 19 | std2 |
| | | **RT** | 31.2 | 50.1 | 88.6 | 108.6 | 173.5 | 248.8 | 258.6 | 419.1 | 464.5 |
| | | **Peak height** | 7.43 | 8.49 | 6.45 | 3.87 | 4.84 | 3.80 | 3.95 | 4.36 | 3.36 |
| | | **Warped RT, Order=2** | 21.06 | 43.90 | 89.49 | 112.67 | 185.59 | 265.72 | 275.80 | 429.28 | 468.75 |
| | | **Warped RT, Order=3** | 19.00 | 43.61 | 91.84 | 115.90 | 189.44 | 266.42 | 275.80 | 410.06 | 441.80 |

**Table S6.** Peak identification performance comparison with fast PTW. The full peak list in $Chrom_1$ is treated as a reference for alignment. PTW warping functions are applied with order 2 and 3. All retention times in the table are provided in seconds. Table **(A)** summarizes the full peak lists (retention times and peak heights) in $Chrom_1$, $Chrom_8$, and $Chrom_9$, as well as warped retention times. The retention time of each compound after warping is close to that in the reference (with a difference of a fraction of a second to a few seconds). Table **(B)** summarizes the subset peak lists in Tests 5 and 7 (generated out of $Chrom_8$), and Tests 10 and 11 (generated out of $Chrom_9$), as well as warped retention times. The retention time of a compound after warping deviates significantly from that in the reference (by a fraction of a second to hundreds of seconds), suggesting that the fast PTW may not be able to handle the situation when only a subset of the target compounds is present in the sample.